\documentclass[%
 reprint,
superscriptaddress,
longbibliography,
showkeys,
 amsmath,amssymb,
 aps,
pre, 
]{revtex4-2}

\usepackage{graphicx}  
\usepackage{dcolumn}   
\usepackage{bm}        
\usepackage{amssymb}   
\usepackage{amsmath}   
\usepackage{mathtools} 
\usepackage{xcolor}
\usepackage{ulem}
\usepackage[colorlinks,linkcolor=blue,urlcolor=blue,citecolor=blue]{hyperref}

\usepackage{dcolumn}
\usepackage{bm}

\usepackage{hyperref}
\hypersetup{
    colorlinks=true,
    linkcolor=blue,
    filecolor=magenta,      
    urlcolor=cyan,
    citecolor = purple,
    pdftitle={Overleaf Example},
    pdfpagemode=FullScreen,
    }
\usepackage{array}
\usepackage{float}

\newcommand{\ee}{\end{equation}} 
\newcommand{\be}{\begin{equation}}

\usepackage[mathscr]{euscript}  
\usepackage{xcolor}  

\usepackage{comment}
\usepackage{float}
\usepackage{xr}
\makeatletter
\newcommand*{\addFileDependency}[1]{
  \typeout{(#1)}
  \@addtofilelist{#1}
  \IfFileExists{#1}{}{\typeout{No file #1.}}
}
\makeatother

\newcommand*{\myexternaldocument}[1]{%
    \externaldocument{#1}%
    \addFileDependency{#1.tex}%
    \addFileDependency{#1.aux}%
}

\myexternaldocument{SI}

\usepackage{soul}

\begin{document} 

\preprint{APS/123-QED}

\title{Thermodynamic Cost of {Recurrent} Erasure}

\author{Deepak Gupta}\thanks{Equal contribution; phydeepak.gupta@gmail.com}
\affiliation{Institut für Theoretische Physik, Technische Universität Berlin, Hardenbergstraße 36, D-10623 Berlin, Germany}

\author{Kristian St\o{}levik Olsen}
\affiliation{Institut für Theoretische Physik II - Weiche Materie, Heinrich-Heine-Universität Düsseldorf, D-40225 Düsseldorf, Germany}

\author{Supriya Krishnamurthy}\thanks{Equal contribution; supriya@fysik.su.se}
\affiliation{Department of Physics, Stockholm University, 106 91, Stockholm, Sweden}

\date{\today}

\begin{abstract}
Recent experiments have implemented resetting by means of a time-varying external harmonic trap whereby the trap stiffness is changed from an initial to a final value in finite-time and then the system is {\it reset} when it relaxes to an equilibrium distribution in the final trap. Such setups are very similar to those studied in the context of the finite-time Landauer erasure principle. We analyze the thermodynamic costs of such a setup by deriving a moment generating function for the work cost of {recurrently changing the trap stiffness in finite-time, thereby maintaining a non-equilibrium steady state.}
We analyze the mean and variance of the work required for a specific experimentally viable protocol and also obtain  
an optimal protocol which minimizes the mean cost. 
For both these procedures, our analysis captures both the large-time and short-time corrections. For the optimal protocol, we obtain a closed form expression for the mean cost for all protocol durations, thereby making contact with earlier work on geometric measures of dissipation-minimizing optimal protocols that implement information erasure.
\end{abstract}

\pacs{Valid PACS appear here} 
\maketitle


\section{Introduction}
\label{intro}

Stochastic resetting  refers to processes in which a system's natural evolution is interrupted and restarted according to some predefined scheme, naturally driving the system out of thermal equilibrium \cite{evans2011diffusion,evans2011diffusion,evans2020stochastic}. Rich behaviors, such as non-trivial non-equilibrium steady state properties \cite{pal2015diffusion,nagar2016diffusion}, anomalous relaxation dynamics \cite{majumdar2015dynamical,gupta2019stochastic,di2025partial} and potential for optimization in search processes \cite{reuveni2016optimal,evans2024stochastic}, have attracted much attention over the past decade, both theoretically and experimentally.
Given this rich behavior and the multitudinous applications, it is very natural to 
consider the thermodynamic cost of  a resetting operation  or the cost associated with maintaining a steady-state resetting process. Particularly interesting are questions relating to fundamental thermodynamic bounds for such costs.

These questions take on an even greater significance
in the light of the fact that
the conceptually simple yet powerful renewal structure built into a resetting process (every reset erases memory and correlations of the system’s past evolution) also hints at deep connections to information erasure \cite{Landauer, Bennett}.  
Early studies on the thermodynamic cost of resetting~\cite{fuchs2016stochastic,busiello2020entropy}, quantified this connection by
bounding the average entropic cost for (instantaneous) resetting in analogy with Landauer's principle of information erasure \cite{Landauer}.
However, these studies did not take into account the actual work needed to implement a reset, which would necessarily also be influenced by the mechanism used to accomplish resetting, finite-time effects, as well as imperfections in the actual resetting. 

Recent experiments \cite{Besga_2020, Faisant_2021, goerlich2024tamingmaxwellsdemonexperimental} that implement resetting  provide a very easy and elegant framework within which to investigate such questions further.
In these experiments, a colloidal particle moves either freely or within a trap until it is reset. The resetting is implemented via a resetting trap generated by optical tweezers. The resetting trap is activated periodically or stochastically and each time it is activated it is left to act for a duration which is long enough for the particle to thermalize. Once thermalization occurs and the particle's distribution has relaxed to the thermal Boltzmann state (characterized by the resetting potential), the particle is considered reset. Note that in this case the particle is not reset to an exact location (an experiment that implements resetting to an exact location is described in \cite{tal_friedman_2020}), but instead to a position drawn from the equilibrium thermal distribution in the resetting trap.

In this guise too,  the resetting problem can be posed as an information erasure problem, whereby, in analogy to how classical bit erasure is implemented via optical traps \cite{Berut2012, Bérut_2013, Bérut_2015, Jun_2014, Gavrilov_2016}, one could ask  what the thermodynamic cost is for turning on the resetting trap, very slowly, instantaneously or in finite-time, and hence erasing all previous information at a (time-dependent) cost. In analogy with Landauer's principle, it is also very interesting to understand what the fundamental bound is on such a cost now taking into account the physical constraints of both time and energy which need to be spent to carry out the erasure.
 
In previous work,  we have studied such a system and obtained the average and fluctuations of the work needed to turn on the resetting trap \cite{olsen2024thermodynamic, olsen2024thermodynamic2} or 
the  heat dissipated \cite{mori2023entropy} as  the particle relaxes in the resetting trap. 
Other studies involving ``first-passage-resetting", i.e., when an optical trap is kept switched on until the particle makes a first-passage to a specified location, have also considered average work costs~\cite{Singh_2024, gupta2022work,pal2023thermodynamic}
and fluctuations~\cite{gupta2022work}.
In all these studies however,  the resetting trap is considered to be switched on instantaneously.
If however, in analogy with several studies on finite-time bit erasure \cite{Berut2012, Bérut_2013, Zulkowski_2014,  Bérut_2015, Jun_2014,Gavrilov_2016, boyd2018shortcutsthermodynamiccomputingcost, Saira_2020, Karel_2020}, one is interested in  lower bounds on the thermodynamic cost of resetting, it is necessary to 
consider the cost  as a function of the time it takes to switch on the resetting trap.
Such a cost will now also depend on the {\it protocol} used for turning on the trap, namely on how the trap parameters are modified to transform the initial trap into the final one.

If the state of the system at the start 
is an equilibrium state, 
the second law of thermodynamics places a bound on the minimum work required to change the control parameter; this is just the equilibrium free energy difference between the initial and final state.  
Any protocol that pays only this minimum cost is a protocol that operates quasi-statically, so that the system is never out of equilibrium for the entire duration. 
In the context of bit erasure, many experiments \cite{Berut2012, Bérut_2013, Bérut_2015, Jun_2014,Gavrilov_2016,Saira_2020} have been performed to probe both the bound as well as
the actual cost to carry out the erasure process in finite time. Several theoretical works \cite{Zulkowski_2014,boyd2018shortcutsthermodynamiccomputingcost, Karel_2020,Ludo_2023} also address this issue.

Although it is possible to look at specific, experimentally viable protocols, theoretically, such questions are best addressed
in the context of {\it optimal} protocols; a protocol engineered so as to minimize some cost of interest such as the mean work performed or the mean dissipation.
Two optimization problems that are typically studied in this context are: 1) Designing optimal protocols that transition between two specified distributions within finite time \cite{Aurell_2011, 
aurell2012refined,Zhang_2019, Dechant_2019, Chen_2020, Nakazato_2021, Grant_2023} and 2) designing optimum protocols that minimize the cost needed to shift between two different potential energy landscapes, often harmonic, in finite time \cite{schmiedl2007optimal, Schmiedl_2008, Sivak_2012, plata2019optimal, Chen_2020, Ito_2021, Abiuso_2022, Zhong_2022}. A different but related class of problems, inspired by the so-called shortcut to adiabatically processes \cite{Muga_2010_1,Muga_2010_2, Schaff_2010, Schaff_2011, Muga_review_2019}, study optimal protocols devised so as to take the system from one equilibrium state to another in a time much faster than the intrinsic relaxation time of the system \cite{martinez2016engineered}.

In this paper, we address the second optimization problem mentioned in the above paragraph. Namely, we address the question of estimating the work cost of a classical overdamped system when changing the potential from $U$ to $V$, effectively implementing an erasure in a fixed time $\tau$. We formally write down the moment generating function for this process for \textit{any} $U$ and $V$. However, we carry out explicit calculations for the mean and variance of the work for harmonic potentials where we transform a harmonic potential $U$ to a harmonic potential $V$ by changing the stiffness of the trap in a finite time $\tau$ according to a generic experimentally viable protocol as well as an optimal protocol.

\begin{figure}
    \centering
    \includegraphics[width = \columnwidth]{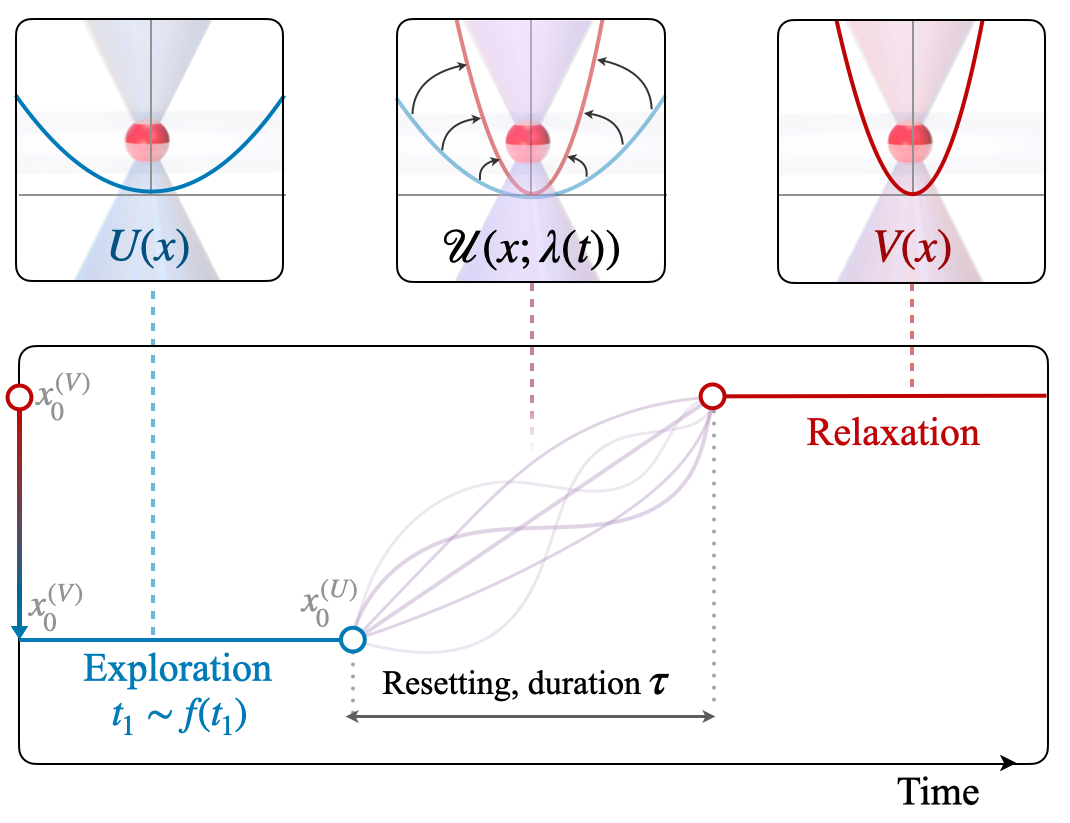}
    \caption{  Schematic for an erasing cycle. $V(x)$: Resetting potential. $U(x)$: Exploration potential. $\mathcal{U}(x;\lambda(t))$: Potential with a time-dependent protocol $\lambda(t)$ such that $\mathcal{U}(x;\lambda(0)) = U(x)$ and $\mathcal{U}(x;\lambda(\tau)) = V(x)$. The exploration time-interval $t_1$ is drawn from a distribution $f(t)$ and duration for the protocol is $\tau$. $x_0^{(V)}$ and $x_0^{(U)}$, respectively, are the particle's positions at the beginning and end of the exploration phase. 
    }
    \label{fig:scheme}
\end{figure}

This system has been studied very extensively in the past, in
experiments \cite{Carberry_2004, Blickle2011RealizationOA, martinez2016brownian, martinez2016engineered, ESE2_2018, Genet_2020, Besga_2020, goerlich2024tamingmaxwellsdemonexperimental} and theoretically \cite{Speck_2004,schmiedl2007optimal,Schmiedl_2008, Engel_2009, Minh_2009, Speck_2011, Nickelsen_2011, Abreu_2011, Kwon_2013, Ryabov_2013, manikandan2017asymptotics,Solon_2018, manikandan2018exact}. 
In contrast to most of these previous studies (two exceptions are  \cite{goerlich2024tamingmaxwellsdemonexperimental} which studies this case albeit from a different aspect and \cite{Abreu_2011} who study an out-of-equilibrium though Gaussian state created by a measurement), we consider 
a steady-state system obtained by repeated applications of our erasure scheme shown in Fig.~\ref{fig:scheme}. The resulting state that needs to be erased is hence  an out-of-equilibrium, non-Gaussian state.
 This fact has implications for the work cost of both generic as well as optimal protocols. As noted recently in the context of bit erasure \cite{Lutz_2021}, it is possible, when beginning with an out-of-equilibrium initial state, to get costs lower than $k_{\rm B}T \ln 2$
in agreement with a generalized Landauer bound \cite{Esposito_2011}.
We study this aspect in-depth for our system by formally obtaining, under very general conditions, the full moment generating function of the work performed in one cycle.
From the expression for the moment generating function, we are
able to compute the average work required to 
carry out one erasure cycle (as in Fig. \ref{fig:scheme}) in the case of harmonic traps, both for a generic experimentally-viable protocol as well as optimal protocol. Our results hold for all time and not only in the short or large-time limits as in most earlier works. Our formalism also allows us
in principle to look at and optimize higher moments of the work, or Pareto-optimal  fronts encoding trade-offs (as done theoretically in \cite{Watanabe_2022},  numerically in \cite{Solon_2018}, or in Ref. \cite{rolandi2023optimal} for quantum systems).

The plan of this paper is as follows. In the next section, we present all our main results. The model is presented in 
Section \ref{model} and the derivation of the moment generating function is given in Section \ref{mgf}. The moment generating function gives us access to the mean and fluctuations of the work cost for any protocol. Section \ref{protocol} presents the analysis for a specific protocol where a rich structure emerges in parameter space of different behaviors of the mean work cost. To better understand bounds on the work cost, in Section \ref{sec:OPB}, we present the analysis of an optimal protocol for all protocol durations. Finally in  Section \ref{summary}, we present our take on interesting research directions to pursue.

\section{Results and discussions}
\label{results}

\subsection{Model}
\label{model}
Our system is subjected to a sequence of erasing protocols in time. 
Each erasing cycle (see Fig.~\ref{fig:scheme}) involves a single resetting event composed of two switching processes: 1) An instantaneous jump from the resetting (or stiff) potential $V(x)$ to an exploration (or shallow) potential $U(x)$ at time $t=0$, after which the system stays in the potential $U(x)$ for a time distributed according to some temporal distribution $f(t)$. We call this phase the exploration phase. 2) After the completion of the exploration phase, a time varying potential $\mathcal{U}(x;\lambda(t))$ is switched on, in such a way that at the beginning and end of the resetting protocol, respectively, $\mathcal{U}(x;\lambda(t=0))=U(x)$ and $\mathcal{U}(x;\lambda(t=\tau))=V(x)$. Here, $\tau$ is the duration of the potential-switching time. We call this phase the resetting phase.  Once the potential $V(x)$ has been turned on, the system stays a fixed amount of time in this potential, say a few times the relaxation-time, so that at the end of this relaxation phase, the system has relaxed  to the Boltzmann distribution in this potential.  At this point the cycle is complete and we switch instantaneously back to $U(x)$. For convenience, henceforth, we drop the explicit mention of the time-dependence from $\lambda(t)$.
The time evolution of the system consists of a series of such erasing cycles.

The system's dynamics is governed by the overdamped Langevin equation:
\begin{align}
    \dfrac{dx}{dt} = -\beta D\dfrac{\partial \mathcal{U}(x;\lambda)}{\partial x} + \sqrt{2D}\, \eta(t)\ ,
\end{align}
for the inverse temperature $\beta \equiv (k_{\rm B}T)^{-1}$, and the diffusion constant $D$. Notice that the above equation is valid for both exploration and relaxation phase by respectively replacing $ \mathcal{U}(x;\lambda)\to U(x)$ and $ \mathcal{U}(x;\lambda)\to V(x)$. In the last term  $\eta(t)$ is a Gaussian thermal white noise which has zero mean and delta-correlations in time: $\langle \eta(t)\eta(t') \rangle = \delta(t-t')$.

In this paper, we are interested in the cost of  performing a complete erasing cycle. By the definition of our erasing cycle, this consists of two contributions.
For the initial instantaneous jump, the performed work along a single stochastic trajectory is the change in the energy due to an instantaneous switch from potential $V(x)$ to $U(x)$,
\begin{align}
    w_1(x) &\equiv U(x) - V(x)\ , \label{w-IJ}
\end{align}
whereas the work performed during the change of the potential from $U(x)$ to $V(x)$ in a time-dependent manner via the control parameter $\lambda(t)$ is
\begin{align}
    w_2[x(\cdot)] &\equiv \int_0^{\tau}~dt~\dot \lambda \dfrac{\partial \mathcal{U}(x;\lambda)}{\partial \lambda}\ ,\label{w-SL}
\end{align}
for the switching time $\tau$. Here, $x(\cdot)$ represents the system's trajectory. Therefore, the total stochastic work for one implementation of the erasing cycle is $w = w_1 + w_2$. Notice that during the relaxation phase, there is no work contribution since the control parameter is fixed.
The stochasticity in this quantity has its origins in two factors: 1) the system's size is small, so thermal fluctuations are significant, and 2) the time to stay in the exploration phase is drawn from a distribution $f(t)$.

\subsection{Moment Generating Function}
\label{mgf}
For a trajectory containing 
  $n$ erasure cycles, i.e., $n$ replicas of the cycle shown in Fig.~\ref{fig:scheme}, the total stochastic  work performed  will just be  the sum of the stochastic works performed during each of these $n$ cycles. Hence, we  turn our attention first to understanding the distribution of work for a single cycle. The
 moment generating function of work $w$ for a single erasure, implemented according to the scheme in Fig.~\ref{fig:scheme}, is defined as \footnote{For $n$ cycles, the moment generating function is $ \langle e^{k w} \rangle^n$} 
\begin{align}
    C(k,\tau) \equiv \langle e^{k w} \rangle\ ,
\end{align}
where the angled brackets indicate the average over thermal fluctuations, initial condition, and stochastic exploration time $t$. Here  $k$ is the conjugate variable with respect to work $w$. This implies
\begin{align}
C(k,\tau)&=\int_0^\infty~dt~f(t)
\int_{-\infty}^{+\infty}~dx_0^{(V)}~P_{{\rm eq},V}(x_0^{(V)}) \nonumber\\
&\times ~\int_{-\infty}^{+\infty}~dx_0^{(U)}~P_U(x_0^{(U)},t|x_0^{(V)}) \nonumber\\
&\times e^{k[U(x_0^{(V)}) - V(x_0^{(V)})]}\underbrace{\big\langle  e^{k w_2} \big\rangle_{0, x_0^{(U)}}}_{C_0(k,\tau;x_0^{(U)})}\ , \label{eq:mgf}
\end{align}
where we have substituted $w_1$ from Eq.~\eqref{w-IJ}. Several comments are in order. $C_0(k,\tau;x_0^{(U)})$ is the moment-generating function of work performed while changing the potential from $U$ to $V$ in a time-dependent manner~\eqref{w-SL}, and herein we average the trajectories emanating from the position $x_0^{(U)}$. This $x_0^{(U)}$ is the position of the particle at the end of the exploration phase, therefore, $x_0^{(U)}\sim P_U(x_0^{(U)},t|x_0^{(V)})$, where $x_0^{(V)}$ is the initial condition for the exploration phase's trajectories. Notice that $x_0^{(V)}$ is drawn from the initial equilibrium distribution with respect to the potential $V$, $P_{{\rm eq},V}(x_0^{(V)})$, since the system initially instantaneously switches from $V$ to $U$ after the relaxation phase. We have weighted Eq.~\eqref{eq:mgf} with respect to a temporal density, $f(t)$, of time intervals in the exploration phase.

\begin{figure}
    \centering
    \includegraphics[width=\columnwidth]{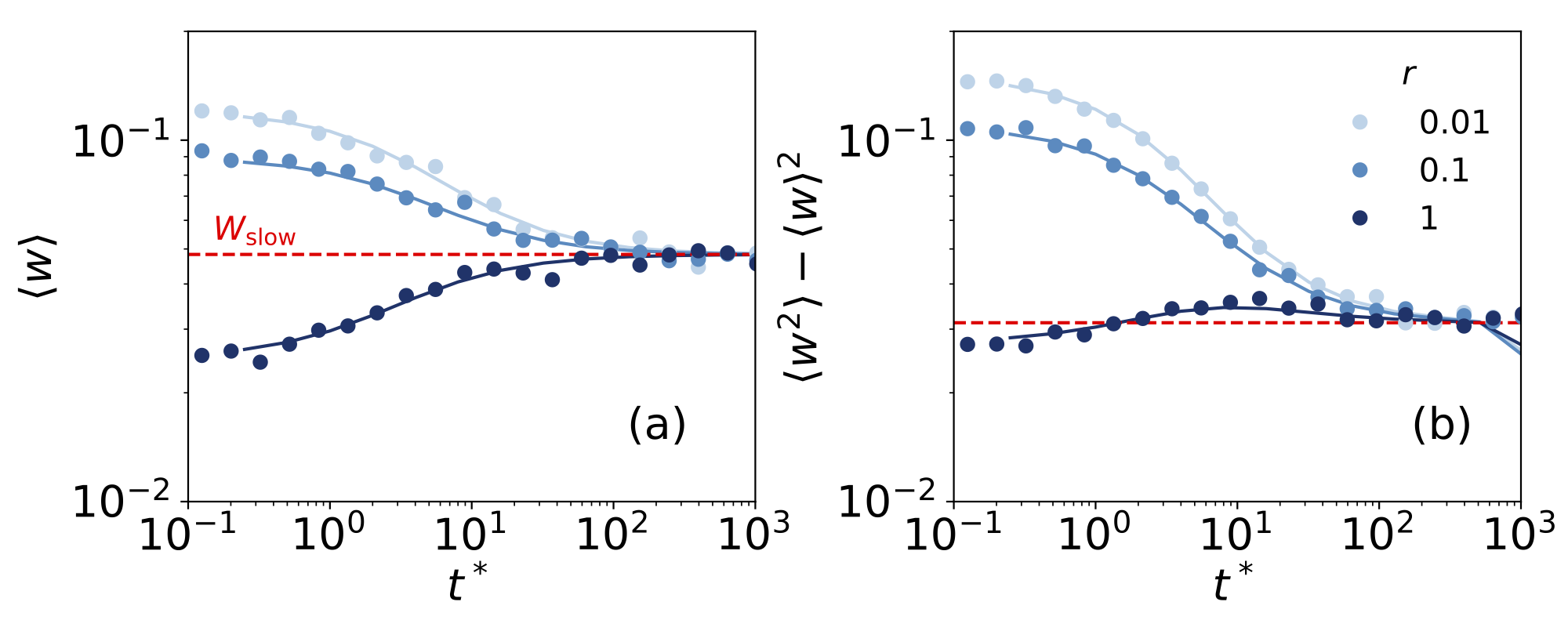}
    \caption{Harmonic exploration and resetting potentials. Mean and variance of work as a function of time $t^*$. Symbols: Numerical simulation. Lines: Analytical results using Eqs.~\eqref{eq:mom-1} and~\eqref{eq:mom2} with $f(t) = r e^{-r t}$ and the protocol in Section \ref{protocol}. {Red horizontal dashed lines indicate asymptotic values of (a) mean work $W_{\rm slow}$~\eqref{eq: w-slow} and (b) variance $[Dt_V(\lambda_V - \lambda_U)]^2/2$ (see Appendix~\ref{var-asym} for more details)}. The color intensity increases with increasing resetting rate $r$. Here, we take the diffusion constant $D=0.5$, $\gamma=1$, $\lambda_U=0.125$, $\lambda_V = 0.25$, and simulation time $\tau = 10~t^*$. Number of resets: $10^5$.}
    \label{fig:mean-var}
    
\end{figure}

Inverting the above equation~\eqref{eq:mgf} for any form of potential seems difficult; nevertheless, it provides the means  to obtaining the $n$th order moments of work. This is done as follows. 
Differentiating both sides of Eq.~\eqref{eq:mgf}  $n$-times with respect to $k$ and setting $k=0$, we get,
\begin{align}
    \langle w^n\rangle &= \int_0^\infty~dt~f(t)~\int_{-\infty}^{+\infty}~dx_0^{(V)}~P_{{\rm eq},V}(x_0^{(V)})\nonumber\\
    &\times~\int_{-\infty}^{+\infty}~dx_0^{(U)}~P_U(x_0^{(U)},t|x_0^{(V)})\nonumber\\
    &\times\sum_{m=0}^n\begin{pmatrix}
    n\\
    m
    \end{pmatrix} w_1^{m}(x_0^{(V)})\langle w_2^{n-m}\rangle_{0, x_0^{(U)}}\ . \label{eq:n-mom}
\end{align}
Analytical calculations of the moments for arbitrary potentials is involved. Therefore, in what follows, we restrict ourselves to the case of harmonic potentials, as these are easily implemented and manipulated in experiments~\cite{Besga_2020,Faisant_2021}.  For this case, we obtain the mean and second moment of the work as follows (see Appendix \ref{sec:cum} for more details):
\begin{align}
  \langle w \rangle \equiv  W&= -\dfrac{k_{\rm B}T}{2}\bigg(1 - \dfrac{t_V}{t_U}\bigg)\nonumber\\
   & + \dfrac{D}{2}\int_0^{\tau}~dt~\dot \lambda \bigg[t_U \zeta^2 G(t,0) + 4 \int_0^t~ds~G(t,s)\bigg]\ , \label{eq:mom-1}\\
    \langle w^2 \rangle &= I_1 + I_2 + I_3\ , \label{eq:mom2}
\end{align}
where 
\begin{align}
\label{eq:gts}
    G(t,s)\equiv e^{-2\beta D\int_{s}^t~dt' \lambda(t')}\ ,
\end{align}
and we have defined a dimensionless length 
\begin{align}
    \zeta\equiv \sqrt{1 - (1- t_V/t_U)\tilde{f}(2/t_U)}\ , \label{eq:len}
\end{align} for the trap relaxation time $t_{V,U}\equiv (\beta D\lambda_{V,U})^{-1}$.
{ Here $\lambda_{V,U}$ are the stiffnesses in the traps $U$ and $V$ respectively and $\tilde{f}(s)=\int_0^\infty~dt~e^{-st}~f(t)$ is the Laplace transform of the exploration time density. Note that
$t_V/t_U\leq \zeta^2\leq 1$ for any distribution $f(t)$.}
$I_{1,2,3}$ in Eq.~\eqref{eq:mom2} are respectively evaluated in Eqs.~\eqref{soln-I1}, \eqref{soln-I2}, and \eqref{soln-I3}. 

The moment generating function for the work performed as the stiffness of the harmonic trap  changes, has been studied in earlier work \cite{Speck_2004, Speck_2011, Kwon_2013, Ryabov_2013,manikandan2018exact} but only when the process begins from an equilibrium initial state. ({ We note that Ref.~\cite{Ryabov_2013} has a closed form expression for the moment generating function for arbitrary initial conditions. They investigate it however only for equilibrium initial conditions.}) In the process we  study here,  the initial state  is in general out-of-equilibrium. It becomes a Boltzmann distribution in the trap $U$ only if the mean exploration time $\langle t_1 \rangle \gg t_U$ (see Fig. \ref{fig:scheme}).

\subsection{An experimentally motivated protocol  with a Poissonian resetting rate}
\label{protocol}

While the above analysis and the expression for the moments [Eqs.~\eqref{eq:mom-1} and \eqref{eq:mom2}] are true for any protocol $\lambda(t)$, we specialize to an experimentally feasible protocol~\cite{goerlich2024tamingmaxwellsdemonexperimental}:
\begin{align}
    \lambda(t) = \lambda_U + (\lambda_V - \lambda_U)\tanh[t/t^*]\ . \label{eq:prot}
\end{align}
Here, $t^*$ controls the speed of the protocol: the smaller the $t^*$ the faster $\lambda_U$ transforms to $\lambda_V$. The total average work $W$ [Eq. \eqref{eq:mom-1}]   is the sum of two terms $ W_1 + W_2 $ ; $W_1$ is the mean contribution of the instantaneous switch at the beginning of the cycle [this is the first term in the Eq.~\eqref{eq:mom-1}] and $W_2$ is the mean contribution of the finite-time switch back to potential $V$ [the second term in Eq.~\eqref{eq:mom-1} which we can now calculate explicitly].  

Figure~\ref{fig:mean-var} shows the comparison of analytical mean and variance of the work $W$ [using Eqs.~\eqref{eq:mom-1} and~\eqref{eq:mom2}] for $f(t) = r e^{-r t}$
as a function of $t^*$, where $r$ is the resetting rate. In both simulations and analytical results, we set the duration of the protocol to $\tau = 10~t^*$ so that at the end of protocol $\lambda(\tau)\approx \lambda_V$, as required.  Both the mean and variance approach their respective stationary values in the quasi-static driving regime $t^*\to \infty$.  From Fig.~\ref{fig:mean-var}a we see, however, that unlike transformations between equilibrium states, the work performed during the quasistatic limit is not necessarily the  minimum.   To investigate this further, we focus only on the average work in what follows
and compare what we get from an instantaneous versus quasi-static protocol.

\begin{figure*}
    \centering
    \includegraphics[width=18cm]{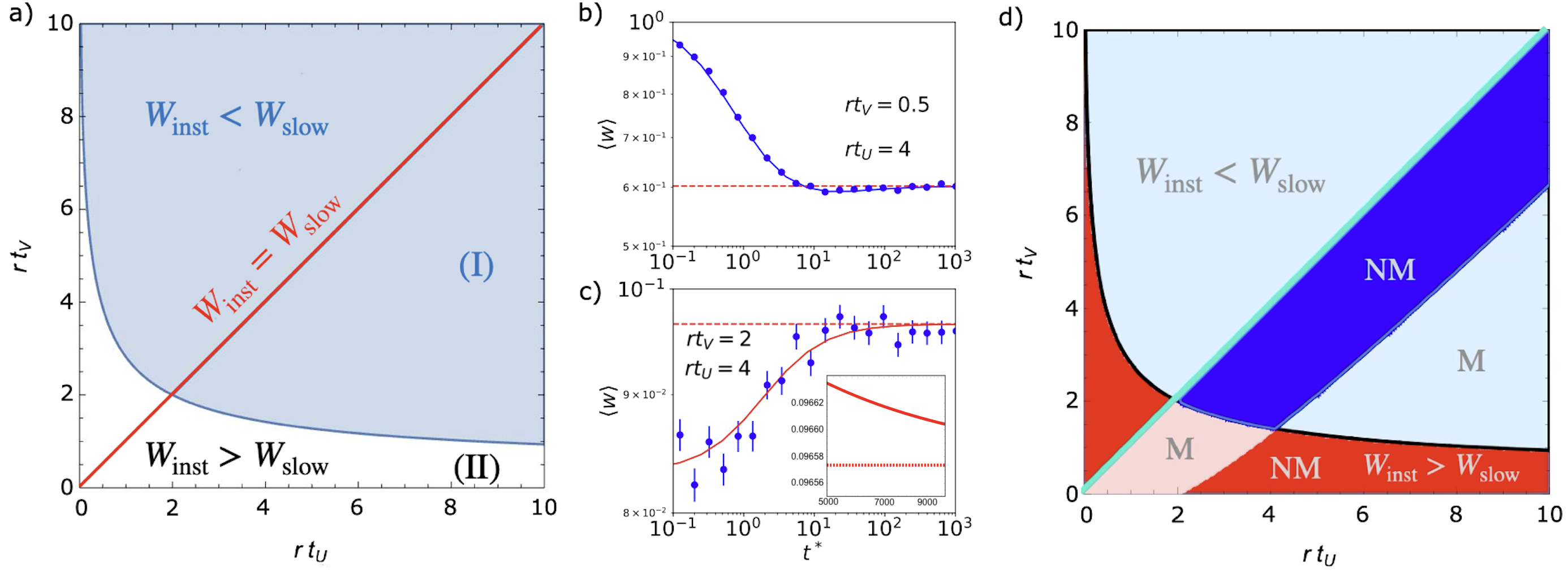}
    \caption{Harmonic exploration and resetting potentials with $f(t) = r e^{-rt}$. a) Phase diagram using Eq.~\eqref{eqn:deltaWc}
    in the $(rt_U,rt_V)$ plane, for resetting rate $r$ and relaxation time $t_{V,U}\equiv (\beta D \lambda_{V,U})^{-1}$. b-c) Mean work as a function of time $t^*$; {inset in panel c) shows work as a function of $t^*$ (where $t^*$ is longer than that of the main panel). Note that the approach to the asymptotic value is from above in (c), indicating non-monotonic behavior at large times.}   Lines: Analytical results Eq.~\eqref{eq:mom-1}. The horizontal dashed line indicates $W_{\rm slow}$~\eqref{eqn:deltaWb}. Here, we take the diffusion constant $D=1$. Number of resets: $10^5$. Error bars show one standard error of mean. d) Phase diagram  from panel a) with added panels [obtained using Eq.~\eqref{W2-longtime}] indicating regions where mean work $W$ is non-monotonic ``NM'' or monotonic ``M'' as a function of time.  Blue regions: $W_{\rm inst}<W_{\rm slow}$. Red regions: $W_{\rm inst}>W_{\rm slow}$. }
    \label{fig:combined_phasediag}
\end{figure*}

In the limit $t^*\to 0$ (i.e., the fast switching limit), the total work simplifies to the following two contributions: 1) an instantaneous switch from $V$ to $U$ starting from equilibrium $P_{{\rm eq},V}(x)$, and 2) another instantaneous switch from $U$ to $V$:
\begin{align}
    W_{\rm inst} \equiv \langle [U(x) - V(x)] \rangle_{P_{{\rm eq},V}(x)}  + {\langle}[V(x) - U(x)] \rangle_{P_{{\rm ta},U}(x)}\ , \label{eq:winst}
\end{align}
 starting from the time-averaged distribution:
 \begin{align}
 P_{{\rm ta},U}(x) \equiv \int_0^\infty~dt~f(t)~P_U(x,t) \ ,\label{eq:noneqstate2}
 \end{align}
 in the $U(x)$ potential, where $P_U(x,t) \equiv \int_{-\infty}^{+\infty}dx_0 P_U(x,t|x_0)P^{\rm eq}_V(x_0)$. 
This time-averaged distribution is the non-equilibrium state at the beginning of our resetting/erasure protocol. Note that because of the renewal structure of our scheme, this distribution does not depend on any details of $\lambda(t)$.

In contrast, in the {infinitely} slow switching of potential {{$t^* \to \infty$ ({i.e.}, $t^* \gg t_{U,V}$})}, we expect the average work to be
\begin{align}
    W_{\rm slow} \equiv \langle [U(x) - V(x)] \rangle_{P_{{\rm eq},V}(x)}+ \Delta F^{\rm eq}_{U\to V}\ ,\label{eq: w-slow}
\end{align}
i.e., we expect that changing potential $U$ to potential $V$ sufficiently slowly will give rise to a work cost equal to the free energy difference between the two potentials. For this protocol, we  can explicitly carry out an expansion in the long-time limit (see Appendix \ref{sec:long-time}) and show that this is indeed the case. As we will see in the next section however, this need not necessarily be the case for arbitrary protocols, i.e., an arbitrary protocol connecting out-of-equilibrium states need not have a work cost that equals the equilibrium free energy difference at long times. 

The right-hand side of Eq.~\eqref{eq: w-slow} can also be rewritten in the form of a KL-divergence~\footnote{$D_{\rm KL} [{p(x)||q(x)}]\equiv \int~dx~p(x)\ln\frac{p(x)}{q(x)}$}:
\begin{align}
   W_{\rm slow} =  \beta^{-1} D_{\rm KL}[P_{{\rm eq},V}(x) || P_{{\rm eq},U}(x)]\ , \label{eq:w-slowkl}
\end{align}
for the equilibrium distributions $P_{{\rm eq},V}(x)$ and $P_{{\rm eq},U}(x)$, respectively, with respect to potentials $V$ and $U$. Thanks to the non-negativity of the KL-divergence we expect $W_{\rm slow}\geq 0$. Using Eq.~\eqref{eq:winst} and Eq.~\eqref{eq:w-slowkl}, we get
\begin{align}
     W_{\rm inst} &- W_{\rm slow}= {\langle}[V(x) - U(x)] \rangle_{P_{\rm ta}(x)} - \Delta F^{\rm eq}_{U\to V}\\
     &=\beta^{-1}\int_0^\infty~dt~f(t)~ \bigg[D_{\rm KL}[P_U(x,t)||P^{\rm eq}_{V}(x)]
     \ +\nonumber\\
     &-D_{\rm KL}[P_U(x,t)||P^{\rm eq}_{U}(x)]\bigg]\ . \label{eq-diff-2}
\end{align} 
Thus, the sign of the left-hand side depends on the KL-distance of $P_U(x,t)$ from the equilibrium distributions in the $V$ and $U$ potentials.

For harmonic potentials, we can explicitly compute the total average work in the instantaneous~\eqref{eq:winst} and slow 
switching~\eqref{eq: w-slow} limits, and hence also their difference. The total average work is $W_1+W_2$, where 
\begin{align}
    W_1 =  \dfrac{k_{\rm B}T}{2}\bigg(\dfrac{t_V}{t_U}-1\bigg)\ , 
\end{align}
and 
\begin{subequations}
\begin{align}
        W_{2,\rm inst} &=  \dfrac{k_{\rm B}T}{2}\bigg(\dfrac{t_U}{t_V}-1\bigg)\zeta^2\ , \label{eqn:deltaWa}\\
        W_{2, \rm slow} &= \Delta F^{\rm eq}_{U\to V} = \dfrac{k_{\rm B}T}{2}\ln \dfrac{t_U}{t_V} \ ,\label{eqn:deltaWb}\\
        W_{\rm inst} - W_{\rm slow} &=\dfrac{k_{\rm B}T}{2}\bigg[\bigg(\dfrac{t_U}{t_V}-1\bigg)\zeta^2 -  \ln \dfrac{t_U}{t_V} \bigg]\ , \label{eqn:deltaWc}
\end{align}
\end{subequations}
where the dimensionless length $\zeta$ is defined in Eq.~\eqref{eq:len}. 


Figure~\ref{fig:combined_phasediag}a shows the parameter regimes for $f(t) = r e^{-rt}$, where the average work $W_{\rm inst}$  [Eq.~\eqref{eqn:deltaWa}] is larger or smaller than $W_{\rm slow}$ [Eq.~\eqref{eqn:deltaWb}]; 
in the unshaded region $W_{\rm inst} > W_{\rm slow}$ otherwise $W_{\rm inst} < W_{\rm slow}$. 
Note that our erasure protocol can lie anywhere below the red diagonal (where $t_V < t_U$). In the shaded region, switching the potential in a finite time costs less work than switching quasi-statically, and is therefore a finite-time cost-effective region.

Figure~\ref{fig:combined_phasediag}b-c shows the mean total work $W$  (in the region below the diagonal line in Fig.~\ref{fig:combined_phasediag}a) as $t_V$ increases from region II to I for a fixed $t_U$ for $f(t) = r e^{-rt}$. We can see that the  work can be made smaller for smaller $t^*$ in region I of phase diagram~\ref{fig:combined_phasediag}a. Additionally, { Figs.~\ref{fig:combined_phasediag}b-c} exhibit the non-monotonic nature of the average work for some values of the parameters. To understand where in parameter space  this occurs, we need to  explore the
large but finite $t^*$ to understand how the $t^* \rightarrow \infty$ limit is approached.

In order to do this, we adapt a technique sketched in \cite{Kwon_2013} to our system and  expand the average work $W_2$ in the slow-switching limit ($t^*\to \infty$) to first order in $ 1/t^*$(see Appendix~\ref{sec:long-time} for a detailed derivation). We show  that the difference of average work $W_2$ from the equilibrium free-energy difference in this limit is given by the correction term
\begin{align}
    W_2 - \Delta F^{\rm eq}_{U\to V}  
     = \alpha \dfrac{k_{\rm B}T}{4} t_U\bigg[{\zeta^2} -\dfrac{1}{2}\left(1+\frac{t_V^2}{t_U^2}\right)\bigg] + \mathcal{O}(\alpha^2) \ ,\label{W2-longtime}
\end{align}
where $\alpha\propto 1/t{^*}$~\eqref{alpha:cal} is the slowness-parameter in the limit $t^*\to \infty$. 
Depending on the sign of the term inside the square brackets in Eq.~\eqref{W2-longtime}, the approach to
$0$ is  either positive or negative. Note that beginning from an equilibrium state (as in \cite{Kwon_2013} which we can also get by putting $\zeta^2 =1$)
results in a purely positive correction term. The fact that we get either a positive or negative contribution from the correction term is purely a result of the non-equilibrium state at the start of the protocol.
Equation~\eqref{W2-longtime} together with the phase diagram in Fig.~\ref{fig:combined_phasediag}a, hence helps us identify regions where the average work is  non-monotonic as a function of $t^*$. This is displayed as a modified phase diagram in Fig.~\ref{fig:combined_phasediag}d.  

\begin{figure}[t!]
    \centering
    \includegraphics[width=\columnwidth]{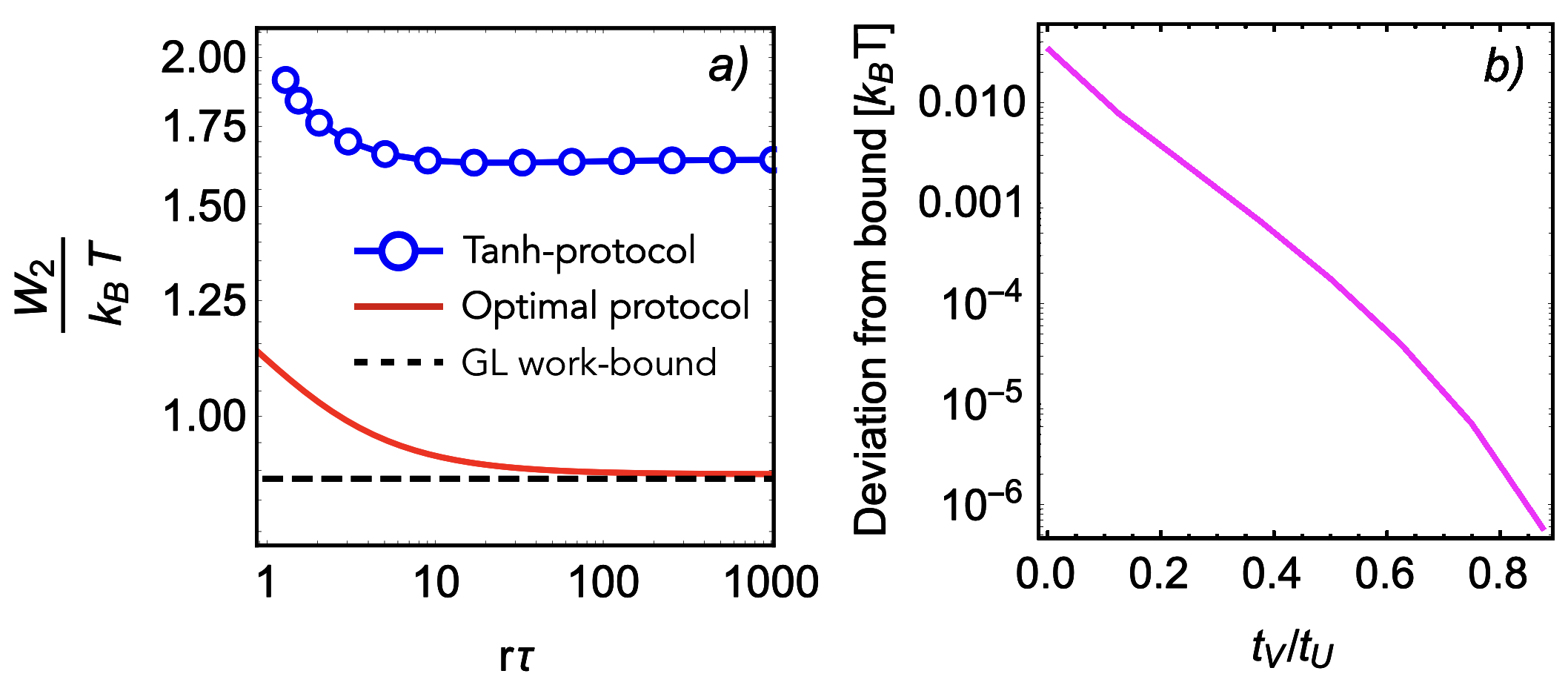}
    \caption{a) Mean work $W_2$ as function of (dimensionless) protocol duration for the {hyperbolic tangent protocol~\eqref{eq:mom-1}} (points) and the {optimal protocol Eq.~\eqref{eq:work_opt_2}} (solid). We use $f(t) = r e^{-rt}$.
    The dashed line represents the theoretical bound~\eqref{KLbound-main}. Here $rt_U = 4, rt_V = 0.5$. b) Deviation from the theoretical GL work-bound~\eqref{KLbound-main}  in the quasi-static limit as a function of $t_V/t_U$. Parameters used are $D =\gamma = k_{\rm B}T = r = 1$, and $t_U =4$. }
    \label{fig:optimalvstanh}
\end{figure}
\subsection{Bounds on the average total work $W$}
\label{sec:OPB}
\subsubsection{ Long-time limit}
Even though the above results hold for a specific protocol, it is clear that the equilibrium free energy difference is no longer in general a lower bound on the average work required for performing our erasure cycle, when the system starts from an out-of-equilibrium state (as was also noticed in the context of bit erasure in \cite{Lutz_2021}). 
It is interesting to understand nevertheless if there are bounds on the total average work even in this case.
As we have seen, the 
total average work is composed of two contributions: a contribution due to the average instantaneous work $W_1$ 
and the average work $W_2$ related to our protocol $\lambda (t)$ for manipulating the trap stiffness. Since the first contribution to the total work does not depend on the protocol at all, we concentrate henceforth on $W_2$. On very general grounds [irrespective of protocol $\lambda(t)$, exploration time distribution $f(t)$, and arbitrary $U(x)$ and $V(x)$], it is expected that a bound on $W_2$ will relate to the {\it non-equilibrium free energy}  differences between the states at the start and end of the protocol.  The minimal work required to transform one out-of equilibrium state into another is  studied in Ref.~\cite{Esposito_2011} and translates in our case to the following expression:
\begin{align}
    W_2 \geq \Delta F^{\rm eq}_{U\to V} +  k_{\rm B}T D_{\rm KL}[P(x,\tau)||P^{\rm eq}_{V}(x)] \  \nonumber \\ 
     -~k_{\rm B}T D_{\rm KL}[P_{{\rm ta},U}(x)||P^{\rm eq}_{U}(x)]\  , \label{KLbound-main} 
\end{align}
where $P_{{\rm ta},U}(x)$~\eqref{eq:noneqstate2} and $P(x,\tau)$, respectively, are the probability of the particle's position at the beginning [$\lambda(0) = \lambda_U$] and end of the protocol [$\lambda(\tau)\approx \lambda_V$].
We refer to this bound as the Generalized Landauer \cite{Esposito_2011} work-bound (GL work-bound). For completeness we  follow Ref.~\cite{Esposito_2011}  in Appendix \ref{GLbound} to show how Eq.~\eqref{KLbound-main} can be derived just from the requirement that the the total entropy production be non-negative for the above process.

Figure~\ref{fig:optimalvstanh}a shows that the GL work-bound~\eqref{KLbound-main} clearly bounds the average work cost $W_2$   of implementing the protocol~\eqref{eq:prot}.
Tighter bounds may be found by minimizing $W_2$ as a functional of the protocol $\lambda(t)$ to obtain an {\it optimal protocol} as discussed below.

\subsubsection{Optimal protocol: long-time bounds}

We search for the optimal protocol of varying the control parameter $\lambda(t)$ that minimizes the average work $W = W_1 + W_2$. Since $W_1$ does not depend on the protocol $\lambda(t)$, it suffices to consider $W_2 $.
Optimal protocols for  harmonic traps have been studied in a wide range of contexts, including overdamped and underdamped Brownian motion \cite{schmiedl2007optimal,gomez2008optimal,Abreu_2011, Aurell_2011, aurell2012refined, Solon_2018,Nakazato_2021,Zhong_2022,Gupta2022}, with constraints on maximum trap stiffness \cite{plata2019optimal,chatterjee2025optimalconstrainedcontrolgenerally} as well as for active particles \cite{davis2024active, DG_OP}. 

In the context we are interested in, the calculation of the optimal (minimal work) protocol can be performed by using the standard Euler-Lagrange minimization of the work functional $W_2[\lambda(t)]$ as carried out in Ref. \cite{schmiedl2007optimal}. As opposed to all earlier studies however, we begin with a non-Gaussian out-of-equilibrium state $P_{{\rm ta}, U}(x_0)$~\eqref{eq:noneqstate2} at the start of the protocol, and this has to be incorporated into the minimization procedure. 
 In Ref. \cite{Abreu_2011}, the authors consider an initial state which has been rendered out-of-equilibrium by a measurement process. The state is however still Gaussian.
 
The optimal work performed on the system is
\begin{align}
    W_2^{\rm opt} 
    &= \frac{k_{\rm B}T}{2} \bigg[(1+c_2 \tau)^2\zeta^2t_U/t_V -\zeta^2  \ + \nonumber\\ 
    &+ 2(c_2\tau)^2\zeta^2t_U/\tau - 2\ln [1+c_2 \tau]\bigg]\ , \label{eq:work_opt_2}
\end{align}
where the dimensionless length $\zeta$ is defined in Eq.~\eqref{eq:len}, and the constant $c_2$ is given by  
\begin{align}
\label{eq:c2m}
     c_2 &= \frac{\sqrt{\frac{\tau}{t_U\zeta^2}\left(\tau/t_V + 2\right)+1}-\left(\tau/t_V+1\right)}{\tau \left(\tau/t_V + 2\right)}\ . 
\end{align}

A detailed derivation of the optimal protocol is included in Appendix~\ref{app:optimal}.  The out-of-equilibrium initial state plays a role in the initial conditions.
In the long time limit ($\tau\to \infty$), we can show that $c_2\tau = -1 + \sqrt{t_V/(t_U\zeta^2)}$, and this gives
\begin{align}\label{eq:slow_optimal2}
    \lim_{\tau\to \infty} W_2^{\rm opt} =
   \Delta F_{U\to V}^\text{eq} + \frac{k_{\rm B} T}{2} \left[\ln\zeta^2 + (1-\zeta^2)\right]\ .
\end{align}
Several points are worth noting. 
Since $\ln\zeta^2 + (1-\zeta^2)$ is  non-positive for all $\zeta$, the long time work $ \lim_{\tau\to \infty} W_2^{\rm opt} \leq F_{U \to V}^\text{eq}$ irrespective of the choice of $f(t)$, with the equality holding only when $\zeta^2 = 1$. Notice that $\zeta^2 = 1$ is only achieved by either $t_V=t_U$ or $\tilde{f}(2/t_U)=0$. The  former is a trivial solution and not interesting.  The latter corresponds to when the system is in equilibrium in the $U$-trap.  Hence, beginning from an out-of-equilibrium state reduces the long-time cost below the equilibrium value.

Equation~\eqref{eq:slow_optimal2} can also be used to get a lower bound on the work as the distribution of the duration in the exploration phase, $f(t)$, is varied. Note first that 
since  $\tilde{f}(2/t_U) \leq 1$, $t_V/t_U \leq \zeta^2 \leq  1$.
In this range,  the function $\ln\zeta^2 + (1-\zeta^2)$ is non-positive and monotonic and attains a maximum value at $\zeta = 1$. This ultimately  gives the bound
\begin{align}\label{eq:wwbound_nonpois}
    \lim_{\tau\to \infty} W_2^{\rm opt} \geq
    \frac{k_{\rm B} T}{2} \left( 1-\frac{t_V}{t_U}\right)\ .
\end{align}
Here, the equality is obtained for 
the case of vanishing duration in the exploration phase, $f(t)=\delta(t)$.
In fact, the same holds for {\it any}  protocol duration $\tau$ in the limit of vanishing exploration duration. This can be seen by expanding $\tilde{f}(2/t_U) = 1- 2 \langle t_1 \rangle/t_U +\dots$ in Eq.~\eqref{eq:work_opt_2}, which results in 
\begin{align}
    \frac{W_2^{\rm opt}}{k_{\rm B} T} =
    \frac{1}{2} \left( 1-\frac{t_V}{t_U}\right) + \left( 1-\frac{t_V}{t_U} \right)^2 \frac{\langle t_1 \rangle}{t_V} + \mathcal{O}(\langle t_1\rangle^2)\ .
\end{align}
The first term on the right-hand side is the negative of the average work done due to the potential switch at time $t=0$, i.e., $-W_1$ [see the first term on the right-hand side of Eq.~\eqref{eq:mom-1}]. Interestingly, to the first order there is no dependence on protocol duration (in the limit of vanishing duration in the exploration phase). Intuitively, in this regime the particle has no time to relax in the shallow trap, and the state at the beginning of the protocol is close to the Boltzmann state in the sharp trap. Hence, the optimal protocol is simply to discontinuously switch almost fully back to the sharp trap and remain there for the rest of the protocol duration, and therefore, the total work, $W_1 + W_2^{\rm opt}$, performed in this limit vanishes. 

In the opposite limit $\zeta^2 = 1 $, the particle has time to equilibrate in the exploration phase, and the protocol connects two equilibrium states. In this case, we expect the work in the quasistatic limit to be given by the free energy difference $\lim_{\zeta^2 \to 1} \lim_{\tau\to \infty} W_2^* = \Delta F_{U\to V}^\text{eq}$, as we verify from Eq.~\eqref{eq:slow_optimal2}. One can also verify that the two limits commute, i.e. $\lim_{\zeta^2 \to 1} \lim_{\tau\to \infty} W_2^* = \lim_{\tau\to \infty} \lim_{ \zeta^2\to 1}  W_2^*$.

{
Another interesting aspect of the optimal protocol is that 
the average work~is monotonic as a function of the duration $\tau$, in contrast to the hyperbolic tangent protocol, i.e., for an optimal protocol, large $\tau$ implies lower mean work. 
This is seen in Fig.~\ref{fig:optimalvstanh}a, where the solid line shows the mean work associated with the optimal protocol.  We also see that the optimal case indeed results in a lower mean work as compared to the hyperbolic tangent protocol, as it should.
The late time work in the optimal protocol approaches the bound~\eqref{eq:slow_optimal2}, which is different as explained above, from the GL work-bound~\eqref{KLbound-main}. In 
Fig.~\ref{fig:optimalvstanh}b we quantify this by plotting the difference between the optimal work bound that we obtain and the GL bound Eq.~\eqref{KLbound-main}. 
The bound is approached monotonically as $t_V \to t_U$ which is the limit corresponding to the trivial case when both potentials are identical and hence both the mean optimal work and the GL work-bound are zero. }

\subsubsection{Information-geometric interpretation}
\label{info-geo}
\begin{figure}[t!]
    \centering
    \includegraphics[width=5.5cm]{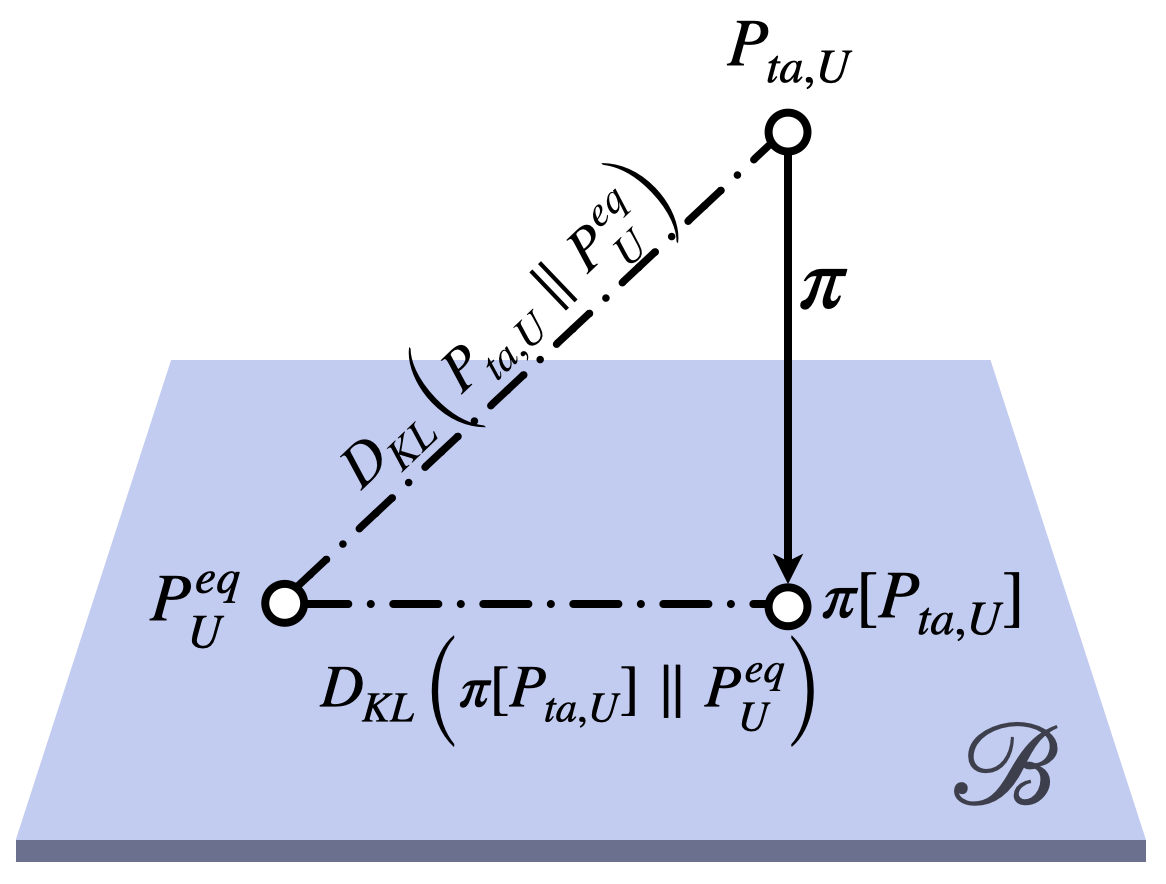}
    \caption{Projection of the true state at the beginning of the protocol into a subspace of Boltzmann densities compatible with the trap. Only the deviation from the initial equilibrium state measured within this subspace can be converted to work. }
    \label{fig:projection}
\end{figure}

Equation~\eqref{eq:slow_optimal2} has a natural information-geometric interpretation. Since our initial state deviates from the Boltzmann state, we expect the information content of this non-equilibrium state to contribute to the work 
\cite{Esposito_2011,Parrondo_2015}. However, the full information content may not always be converted into (negative) work since we are constrained to harmonic trap shapes \cite{kolchinsky2021work}. To make this argument more precise, we first consider the so-called m-projection
\begin{equation}
    \pi[P_{\text{ta},U}](x) = \underset{q\in \mathscr{B}}{\text{argmin}} \:D_\text{KL}( P_{\text{ta},U} \parallel q)\ ,
\end{equation}
into a subspace $\mathscr{B}$ of Boltzmann states compatible with our trap, i.e., Gaussian densities in the context of harmonic traps (Fig.~\ref{fig:projection}). 

Since $q(x)$ is a generic Gaussian density, let's say 
it has mean $\mu$ and standard deviation $\sigma$, the KL-divergence we want to minimize can be written as
\begin{equation}
    D_\text{KL}( P_{\text{ta},U} \parallel q) = - S[P_{\text{ta},U}] + \frac{\langle (x-\mu)^2 \rangle_{\text{ta},U}}{2 \sigma^2}+ \ln \sqrt{2\pi \sigma^2}\ ,
\end{equation}
where $S[P_{\text{ta},U}]$ is the Shannon entropy of the true initial state. To minimize this over $(\mu,\sigma)$, we simply take the derivative with respect to these two parameters and solve for stationary points. This results in $\mu=0$ and $\sigma^2 = \langle x^2\rangle_\text{ta}= D t_U \zeta^2$. Hence, the projected state $\pi[P_{\text{ta},U}](x)$ is simply a Gaussian with matching first two moments of $P_{\text{ta},U}$. 

The information between the true initial state and the instantaneous equilibrium state associated with the initial trap is measured by $D_\text{KL}(P_{\text{ta},U} \parallel P_{U}^{\rm eq})$, which in principle can be converted into work as seen in the 
GL work-bound~\eqref{KLbound-main}. However, due to the constraints on the protocol (i.e., harmonic traps) only the KL-divergence $D_\text{KL}(\pi[P_{\text{ta},U}] \parallel  P_{U}^{\rm eq})$  between the \textit{projected} initial state and the instantaneous equilibrium state (the accessible information) can be converted into work; see Fig.~\ref{fig:projection}. Indeed, using well-established formulas for the KL divergence between two Gaussians, we find that Eq.~\eqref{eq:slow_optimal2} can be rewritten as
\begin{equation}
    \lim_{\tau \to \infty} W_2^\text{opt} = \Delta F^\text{eq}_{U\to V} - k_{\rm B}T \: D_\text{KL}(\pi[P_{\text{ta},U}] \parallel  P_{U}^{\rm eq})\ .
    \label{eq:proj}
\end{equation}
Hence, the deviations from the bound given by Eq.~\eqref{KLbound-main} originates in the information loss associated with the projection of $P_{\text{ta},U} \to \pi[P_{\text{ta},U}]$.

\subsubsection{ Optimal protocol: Finite-time results}

{ The manner in which the bound in Eq.~\eqref{eq:slow_optimal2} is reached} can be better understood in the light of recent work
on geometrical measures of optimal minimum-work or minimum-disspation protocols (see \cite {Blaber_2023} for a recent review on some  of these aspects). To understand these results in our context, we first
get the full $\tau$-dependence of the
optimal work $W_2^{\rm opt}$ by separating the time-independent and time-dependent terms in Eq.~\eqref{eq:work_opt_2} 
by substituting $c_2$ from Eq.~\eqref{eq:c2m}:
\begin{align}\label{eq:w2-optimal2}
    \dfrac{W_2}{k_{\rm B}T} = &\beta\Delta F_{U\to V}^\text{eq} + \frac{1}{2} \left[\ln\zeta^2 + (1-\zeta^2)\right]\ \nonumber \\
    -&\ln \left[ \frac{\sqrt{1+\frac{2t_V}{\tau} + \frac{\zeta^2 t_Ut_V}{\tau^2}}}{(1+\frac{2t_V}{\tau})} +\frac{\sqrt{\zeta^2 t_Ut_V}}{\tau (1+\frac{2t_V}{\tau})}\right] \nonumber \\ -& \frac{t_U \zeta^2}{\tau }\left[ \frac{1}{\zeta} \sqrt{\frac{t_V}{t_U}} \frac{\sqrt{1+\frac{2t_V}{\tau} + \frac{\zeta^2 t_Ut_V}{\tau^2}}}{(1+\frac{2t_V}{\tau})} - \frac{(1+ \frac{t_V}{\tau})}{(1+ \frac{2t_V}{\tau})}\right]\ .
\end{align}

The time-independent terms are just the non-equilibrium free energy difference that we obtained in Eq.~\eqref{eq:slow_optimal2}.
From general considerations \cite{Aurell_2011, aurell2012refined, Karel_2020, Chen_2020, Nakazato_2021, Dechant_2019, Blaber_2023}, we expect
the time-dependent terms to be strictly positive for all time and equal to the entropy production in the system due to finite driving speeds. For overdamped systems, the entropy production has been shown to be related to the $L^2$- Wasserstein distance \cite{Zhang_2019, Nakazato_2021, Dechant_2019} between  specified initial and final distributions. As we have emphasized, our problem involves instead changing a potential $U$ to a potential $V$ in a finite time. However, the structure is similar \cite{Zhong_2022, Blaber_2023}.

It is easy to show in certain limits that the time-dependent terms in Eq.~\eqref{eq:w2-optimal2} are indeed related to the $L^2$- Wasserstein distance. If we carry out a large-$\tau$ expansion on the time-dependent terms, it is easy to see that $W_2$ can be written as
\begin{align}
\label {eq:Wass}
  \dfrac{W_2}{k_{\rm B}T} = &\beta\Delta F_{U\to V}^\text{eq} + \frac{1}{2} \left[\ln\zeta^2 + (1-\zeta^2)\right]\  \nonumber \\  &+ \frac{1}{2\tau} \left [ \sqrt{2t_V} - \zeta \sqrt{2t_U} \right]^2\ .
\end{align}

In this limit the $\tau$-dependent term can be expressed in terms of the square of the $L^2$- Wasserstein distance between two Gaussians, where one of the Gaussians is simply the Boltzmann distribution in the $V$-trap while the other is a Gaussian in a modified shallow trap with variance $D\zeta^2 t_U$. 
This is the Gaussian projection of the time-average nonequilibrium state as mentioned in Sec.~\ref{info-geo}.

When $\zeta^2 =1 $ (which is the limit when $P_{{\rm ta},U}(x) \rightarrow {P_{{\rm eq},U}(x)} $), the second term in Eq.~\eqref{eq:Wass} vanishes and the third term simplifies to the square of the $L^2$-Wasserstein distance between the two Boltzmann distributions in the harmonic potentials $U$ and $V$ \cite{Abiuso_2022}. In this limit Eq.~\eqref{eq:Wass} takes on the form expected from the Jarzynski equality  \cite{Jarzynski_1997} which connects in this context, the expected value of the work to the variance \cite{Speck_2004}. It is interesting that Eq.~\eqref{eq:Wass} has this form for any value of $\zeta$ whereas the Jarzynski equality holds strictly when starting from equilibrium.

In the opposite limit of $\zeta^2 = \frac{t_V}{t_U} $ it is easy to see from both Eqs.~\eqref{eq:w2-optimal2} and~\eqref{eq:Wass}, that the $\tau$-dependent terms entirely vanish. This is consistent with the interpretation of this limit as the trivial case of vanishing exploration duration, as also mentioned earlier.

The instantaneous limit $\tau \rightarrow 0$ can also be taken in Eq.~\eqref{eq:w2-optimal2}. In this case one recovers Eq.~\eqref{eqn:deltaWa}. 
Details of the limits as well as the derivation of Eq.~\eqref{eq:w2-optimal2} are given in Appendix \ref{app:optimal}.

\begin{figure}[t!]
    \centering
    \includegraphics[width=8.65cm]{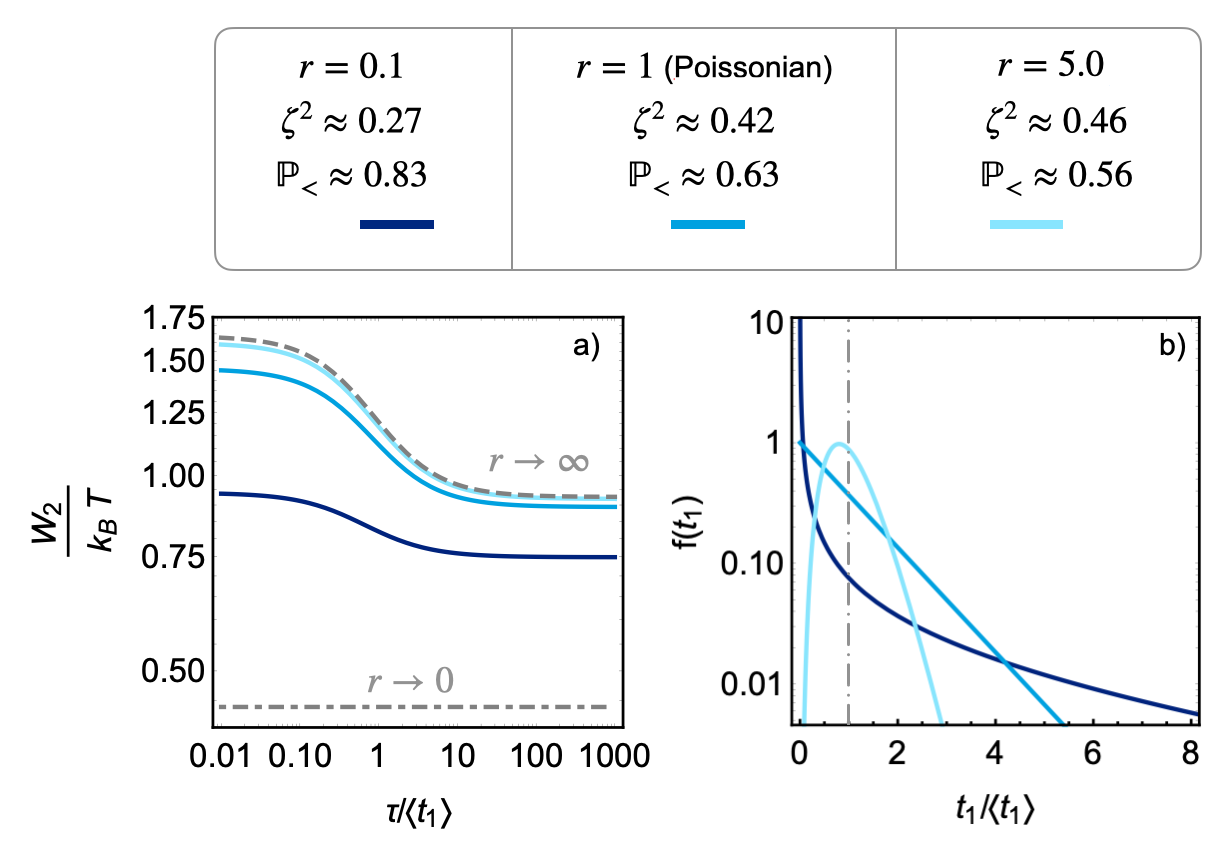}
    \caption{a) Work for a scenario where the exploration durations are drawn from a Gamma distribution with different shapes. The value of $r, \zeta^2$ as well as the sub-mean probability mass $\mathbb{P}_<$ is reported. The dashed line shows the bound when $r  \to \infty$ (small fluctuations around mean), and the dot-dashed line $r  \to 0$ (small fluctuations around zero). b) Corresponding probability density for durations of the exploration phase. Parameters are set to $t_U =4, t_V = 0.5$ and $\langle t_1 \rangle= 1$. }
    \label{fig:nonpoiss}
\end{figure}

 In all of the above, the out-of-equilibrium nature of the initial state is parametrized via $\zeta$ which is itself a function of the strength of potentials $U$ and $V$ as well as the distribution of durations in the exploration phase $f(t_1)$ [Eq.~\eqref{eq:len}].
 To understand the role of $f(t_1)$ better, 
we plot Eq.~\eqref{eq:w2-optimal2} in Fig.~\ref{fig:nonpoiss}a, for a Gamma distribution
\begin{equation}
    f(t_1) =  \frac{r^{r \langle t_1 \rangle}}{\Gamma(r \langle t_1 \rangle)}e^{- r t_1} t_1^{r \langle t_1 \rangle-1} \ ,
\end{equation}
where $r \langle t_1 \rangle \neq 1$ corresponds to deviations from the Poissonian case. In the following, we fix the mean duration in the exploration phase $\langle t_1 \rangle =1 $ and  vary $r$.  This  changes the shape of the distribution $f(t_1)$ (Fig.~\ref{fig:nonpoiss}b). In Fig.~\ref{fig:nonpoiss}a we plot the mean work as a function of protocol duration. We see that, for all protocol durations, 
$W_2$ increases with $r$. To relate this to the shape of 
$f(t_1)$,  we see that $\mathbb{P}_< = \int_0^{\langle t_1 \rangle} dt~f(t)$ grows as $r$ decreases. This implies that large fluctuations which enable  stochastic realizations with sub-mean exploration duration result in a lower work value while suppressing fluctuations in $f(t_1)$ results in higher work values. 
This is in line with our previous discussions on  Eq.~\eqref{eq:wwbound_nonpois}
which showed that the lesser the time in the exploration phase, the lower the value of $W_2$. Indeed, when $r \to 0$ the probability accumulates at zero (Fig.~\ref{fig:nonpoiss}b), resulting in the same work as the bound in Eq.~\eqref{eq:wwbound_nonpois} (dot-dashed line in Fig.~\ref{fig:nonpoiss}a). {When $r\to \infty$ in the Gamma distribution the fluctuations around the mean become vanishingly small, and we approach the limit of deterministic exploration durations. In Fig.~\ref{fig:nonpoiss}a the dashed line shows the resulting work for the corresponding case.}

\section{Discussion and Outlook}
\label{summary}
We have investigated the connections between the thermodynamic cost of an experimental procedure which has been used to implement resetting \cite{Besga_2020,Faisant_2021,goerlich2024tamingmaxwellsdemonexperimental} and earlier well studied problems of information erasure \cite{Esposito_2011, Parrondo_2015} and geometric measures of optimal protocols that minimize work or heat in overdamped stochastic systems  \cite{Aurell_2011, Abiuso_2022, Dechant_2019, Nakazato_2021, Zhang_2019, Karel_2020, Blaber_2023} . The problem we study is very similar to those studied in the above contexts, but a few important distinguishing aspects are the 
out-of-equilibrium nature of the system we study, the moment generating function of work from which we can, in principle, obtain all moments of the work for this system,  as well as the
explicit expression we obtain for the optimal work Eq.~\eqref{eq:w2-optimal2} which holds for all protocol durations $\tau$ and not just in the slow and fast limits as often studied.

We see from the expressions for the optimal work that at late times~\eqref{eq:Wass} appears to be the square of a $L^2$-Wasserstein distance between two Gaussians, one of which has a variance $D \zeta^2 t_U $.
The non-dimensional length $\zeta$ is a quantifier of the out-of-equilibrium initial state at the start of the resetting protocol and itself depends on both potentials $U$ and $V$ as well as the waiting time $f(t)$ in a non-trivial manner [Eq.~\eqref{eq:len}]. In addition this length scale quantifies 
the variance of the projected state $\pi[P_{\text{ta},U}](x)$.
This seems to suggest that, at least for some results, 
we can replace our time-averaged non-equilibrium state $ P_{\text{ta},U}$
by a Gaussian with the same mean and variance,  if all manipulations are only done via harmonic traps. It would be interesting to see how general this result is and whether it translates to other non-harmonic potentials.

Investigating optimal protocols further in the context of the erasure of out-of-equilibrium states is a very interesting direction to pursue. In particular, it would be very illuminating to understand if there are trade-offs between minimizing the mean work and minimizing the variance \cite{Solon_2018, Watanabe_2022} of the work done or the heat dissipated and if this results in phase-transitions in protocol space \cite{Solon_2018}.  It would also be interesting to 
understand if there are cases when these protocols can  be non-monotonic as  
observed in \cite{Zhong_2022}. Investigating the so-called  thermodynamic  metric structure of the optimal-protocol-parameter space \cite{Sivak_2012,Neha_2022,Ito_2023, Grant_2023}  is yet another interesting direction to pursue, as is also the study of optimal transport in discrete cases \cite{Vanvu_2023}.

Finally, coming back to the context of resetting, it is very interesting to also understand thermodynamic costs when there is an absorbing barrier. 
Such an analysis has been done in \cite{pal2023thermodynamic}, though 
resetting has been implemented there by considering a first passage excursion to a specific point in a trap (``first-passage resetting"), unlike in our case when resetting is accomplished when the Boltzmann distribution in the trap has been  reached. 
It would be very interesting to carry out a similar analysis 
as done in \cite{pal2023thermodynamic} for our case. The nature of optimal protocols in the presence of absorbing barriers is also  very interesting to understand
~\cite{pal2023thermodynamic,Singh_2024}.

\section{Acknowledgment}
All the authors are very grateful to Satya Majumdar and Sergio Ciliberto for inspiring discussions on this problem during the Nordita conference ``Measuring and Manipulating Non-Equilibrium Systems” in Stockholm (14-24, October, 2024). SK would also like to thank Patrizia Vignolo for an interesting discussion during the Indo-French workshop on  ``Classical and quantum dynamics in out of equilibrium systems", held at ICTS, Bengaluru, Dec 16th-20th, 2024.
DG gratefully acknowledges the financial support provided by NORDITA and Stockholm University, as well as the support from the Special Research Fund (BOF) of Hasselt University (BOF number: BOF24KV10) during the academic visit in May–June 2024, which facilitated this collaboration.
DG and KSO acknowledge support from the Alexander von Humboldt foundation. SK acknowledges the support of the Swedish Research  Council through the grant 2021-05070.
\appendix


\setcounter{equation}{0}
\renewcommand{\theequation}{\thesection\arabic{equation}}
\setcounter{figure}{0}
\renewcommand{\thefigure}{\thesection\arabic{figure}}
\setcounter{table}{0}
\renewcommand{\thetable}{\thesection\arabic{table}}
\begin{widetext}
\section{Cumulants of the work}
\label{sec:cum}

\subsection{First cumulant}
\label{sec:fcum}
In this section, we provide the detailed calculation of the computation of the first moment [i.e., $n=1$ in Eq.~\eqref{eq:n-mom}] of the work: 
\begin{align}
    \langle w\rangle = \langle w_1\rangle + \langle w_2 \rangle\ . \label{eq:mean-work}
\end{align}

Since the erasing protocol involves an instantaneous jump at time $t=0$, the average work $W_1\equiv \langle w_1\rangle$ is computed by performing the average of Eq.~\eqref{w-IJ} over the initial equilibrium distribution $P_{{\rm eq},V}(x)$, and it gives:
\begin{align}
    W_1 = \int_{-\infty}^{+\infty}~dx~P_{{\rm eq},V}(x)[U(x) - V(x)]\ . \label{eq:avg}
\end{align}

When both resetting and the exploration potentials are harmonic with distinct stiffnesses $\lambda_V$ (stiff-trap) and $\lambda_U$ (shallow-trap), the above integral~\eqref{eq:avg} can be evaluated easily:
\begin{align}
    W_1  = 
    \dfrac{\lambda_U-\lambda_V}{2} D t_V\ , \label{w1avg}
\end{align}
for the stiff-trap's relaxation time $t_V\equiv (\beta D\lambda_V)^{-1}$.

The average work $W_2\equiv \langle w_2\rangle$ can be computed as follows. First, we average the right-hand side of Eq.~\eqref{w-SL} over the thermal noise's history, i.e, an ensemble of trajectories for a fixed initial condition $x_0^{(U)}$:
\begin{align}
    \langle w_2\rangle_{0,x_0^{(U)}} = \int_0^{\tau}~dt~\dfrac{\dot \lambda}{2} \langle x^2\rangle_{0,x_0^{(U)}}\ , \label{eq:w2-avg}
\end{align}
where $x_0^{(U)}$ is the position at the beginning of the resetting phase, and the subscript ``0'' denotes that this is an
average over trajectories only during the interval 
where the control parameter changes.
The position's second moment in the integrand~\eqref{eq:w2-avg} due to change in the potential $U$ to $V$ (in the resetting phase) is 
\begin{align}
    \langle x^2\rangle_{0, x_0^{(U)}} = [x_0^{(U)}]^2 G(t,0) + 2 D \int_0^t~ds~G(t,s)\ , \label{eq:xsq}
\end{align}
for
\begin{align}
\label{eq:gts}
    G(t,s)\equiv e^{-2\beta D\int_{s}^t~dt' \lambda(t')}\ .
\end{align}

Then, the average work $W_2$ in the resetting phase is computed by performing the average of Eq.~\eqref{eq:w2-avg}  over $x_0^{(U)}$ with respect to the time-average position density 
\begin{align}
    P_{{\rm ta},U}(x) \equiv \int_0^\infty~f(t)~\int_{-\infty}^{+\infty}~dx_0~P_{{\rm eq}, V}(x_0)P_U(x,t|x_0)\ ,  \label{app:ta-U-den}
\end{align}
for the resetting time-interval density $f(t)$, the propagator in the exploration phase $P_U(x,t|x_0)$, and equilibrium distribution in the stiff-potential (or the distribution at the beginning of the exploration phase) $P_{{\rm eq}, V}(x_0)$:
\begin{align}
    W_2 =\int_0^\infty~dt''~f(t'')~\int_{-\infty}^{+\infty}~dx_0~P_{{\rm eq},V}(x_0)\int_{-\infty}^{+\infty}~dx_0^{(U)}~P_U(x_0^{(U)},t''|x_0)~\langle w_2\rangle_{0,x_0^{(U)}}\ . \label{eqn:W2}
\end{align}
On the right-hand side, we first notice that
\begin{align}
    P_U(x,t)\equiv \int_{-\infty}^{+\infty}~dx_0~P_{{\rm eq},V}{(x_0)}~P_U(x,t|x_0)=\dfrac{1}{\sqrt{2\pi \sigma_{t}^2}} \exp\bigg[-\dfrac{x^2}{2\sigma_{t}^2}\bigg]\ ,
\end{align}
where the variance in the exploration phase is
\begin{align}
 \sigma_{t}^2\equiv D t_V e^{-2t/t_U} + D t_U (1- e^{-2t/t_U}) \ , \label{sigsq} 
\end{align}
for the $U$-trap's relaxation time $t_U\equiv (\beta D\lambda_U)^{-1}$.

Then, Eq.~\eqref{eqn:W2} becomes:
\begin{align}
    W_2 & =     \int_0^\infty~dt~f(t)~\int_{-\infty}^{+\infty}~dx~P_U(x,t)~\langle w_2\rangle_{0,x}\ . \label{eqn:W2-2}
\end{align}

Using Eq.~\eqref{eq:w2-avg} in Eq.~\eqref{eqn:W2-2}, we get
\begin{align}
    W_2  
     &= \dfrac{D}{2}\int_0^{\tau}~dt~\dot \lambda \bigg[t_U \zeta^2 G(t,0) + 4 \int_0^t~ds~G(t,s)\bigg]\ , \label{eq:w2avg}
\end{align}
where we 
define a  dimensionless length
\begin{align}
    \zeta\equiv \sqrt{1 - (1- t_V/t_U)\tilde{f}(2/t_U)}\ . \label{app:eq:len}
\end{align}

\subsection{Second cumulant}
\label{sec:sec_mom}
{ The variance of the total work can be written as $\langle w^2\rangle - \langle w\rangle^2  = (\langle w_1^2 \rangle - \langle w_1\rangle^2) +  (\langle w_2^2 \rangle - \langle w_2\rangle^2) + 2 (\langle w_1w_2 \rangle - \langle w_1\rangle \langle w_2\rangle)$, where the average work $\langle w\rangle \equiv W$ is discussed in the previous subsection~\ref{sec:fcum}.
Now using Eq.~\eqref{eq:n-mom} we can compute the second moment of the total work $w$.
 }
This will give us three contributions:
\begin{align}
   \langle w^2 \rangle  = I_1 + I_2 + I_3\ , \label{eq:sm}
\end{align}

In the following, we compute each of these contributions for harmonic exploration (shallow) and resetting (stiff) traps.

Let us first compute $I_1$:
\begin{align}
    I_1 &= \int_0^\infty~dt~f(t)~\int_{-\infty}^{+\infty}~dx_0~P_{{\rm eq},V}(x_0)\int_{-\infty}^{+\infty}~dx_0^{(U)}~P_U(x_0^{(U)},t|x_0) w_1^{2}(x_0) =\int_{-\infty}^{+\infty}~dx_0~P_{{\rm eq},V}(x_0)~w_1^{2}(x_0) \\
    & = \dfrac{3}{4} D^2 t_V^2(\lambda_U-\lambda_V)^2\ , \label{soln-I1} 
\end{align}
where we used the definition of $w_1(x)$ given in Eq.~\eqref{w-IJ}.

The computation of $I_2$ is as follows:
\begin{align}
    I_2 &=\int_0^\infty~dt~f(t)~\int_{-\infty}^{+\infty}~dx_0~P_{{\rm eq},V}(x_0)\int_{-\infty}^{+\infty}~dx_0^{(U)}~P_U(x_0^{(U)},t|x_0)\langle w_2^{2}\rangle_{0,x_0^{(U)}} \nonumber\\
    &= \int_0^{\tau}~dt_1\int_0^{\tau}~dt_2~\dfrac{\dot \lambda(t_1)}{2}~\dfrac{\dot \lambda(t_2)}{2}  A(t_1,t_2) \label{soln-I2}\ ,
\end{align}
where $A(t_1,t_2)$ is given in Eq.~\eqref{eq:A} (see Appendix~\ref{sec:Aw2} for more details).
Now let us compute $I_3$. We have
\begin{align}
    I_3 &= 2\int_0^\infty~dt'~f(t')~\int_{-\infty}^{+\infty}~dx_0~P_{{\rm eq},V}(x_0)\int_{-\infty}^{+\infty}~dx_0^{(U)}~P_U(x_0^{(U)},t'|x_0) w_1(x_0)\langle w_2\rangle_{0,x_0^{(U)}}\\
    &=(\lambda_U-\lambda_V)\int_0^\infty~dt'~f(t')\int_{-\infty}^{+\infty}~dx_0~P_{{\rm eq},V}(x_0)~\int_{-\infty}^{+\infty}~dx_0^{(U)}~P_U(x_0^{(U)},t'|x_0)~x_0^2~\int_0^{\tau}~dt~\dfrac{\dot \lambda}{2} \langle x^2\rangle_{0,x_0^{(U)}}\ . \label{I3-pre}
\end{align}

Substituting the propagator of a Brownian particle in the exploration phase at time $t$:
\begin{align}
    P_U(x,t|y) = \dfrac{\exp\bigg[-\dfrac{(x-y~e^{-t/t_U})^2}{2Dt_U(1-e^{-2t/t_U})}\bigg]}{\sqrt{2\pi Dt_U(1-e^{-2t/t_U})}}\ ,
\end{align}
and $\langle x^2\rangle_{0, x_0^{(U)}}$~\eqref{eq:xsq}
on the right-hand side of Eq.~\eqref{I3-pre} and integrating over $x_0^{(U)}$, we get

\begin{align}
    I_3 &=(\lambda_U-\lambda_V)\int_0^{\tau}~dt~\dfrac{\dot \lambda}{2} \int_0^\infty~dt'~f(t')\int_{-\infty}^{+\infty}~dx_0~P_{{\rm eq},V}(x_0)~\nonumber\\
    &\times\bigg[\bigg(x_0^4 e^{-2 t'/t_U} + x_0^2 D t_U(1 - e^{-2 t'/t_U})\bigg)~G(t,0) + 2 D \int_0^t~ds~G(t,s)\bigg]\\
        &=(\lambda_U-\lambda_V)~\int_0^{\tau}~dt~\dfrac{\dot \lambda}{2} \bigg[D^2 t_U t_V \big[1 - \tilde{f}(2/t_U)\big(1-3t_V/t_U\big)\big]~G(t,0)+ 2 D^2t_V \int_0^t~ds~G(t,s)\bigg]\ .\label{soln-I3}
\end{align}

Thus, summing $I_1$~\eqref{soln-I1}, $I_2$~\eqref{soln-I2}, and $I_3$~\eqref{soln-I3}, we get the second moment of work~\eqref{eq:sm}. By subtracting the square of the mean work~\eqref{eq:mean-work}, we can get the second cumulant of the work.

\section{Long-time expansions}
\subsection{Long-time expansion of the average work for the $\tanh$ protocol}
\label{sec:long-time}
The specific protocol we consider, motivated by recent experiments \cite{goerlich2024tamingmaxwellsdemonexperimental} is
\begin{align}
 \lambda(t) = \lambda_U + (\lambda_V-\lambda_U)\tanh[t/t^*]   
\end{align}
In the slowest protocol limit ($t^* \to \infty$), we can write 
\begin{align}
    \lambda(t) \approx \lambda_U + (\lambda_V-\lambda_U)t/t^* = \lambda_U (1 + \alpha t)
\end{align}
\begin{align}
\label{alpha:cal}
    \alpha \equiv \frac{\lambda_V-\lambda_U}{\lambda_U t^*}
\end{align}
In the slowest protocol limit $t^*\to \infty \implies \alpha \to 0$  with the product $\alpha t^* $ fixed so that $ \alpha t^*=\frac{\lambda_V-\lambda_U}{\lambda_U}$. 
Henceforth, we will replace $t^*$ by $\tau$ (since these are the same for a linear protocol). In terms of the variables $t_U\equiv (\beta D\lambda_U)^{-1}$ and 
$t_V\equiv (\beta D\lambda_V)^{-1}$
\begin{align}
\label{alpha:cal1}
    \alpha  \tau = \frac{t_U}{t_V} -1
\end{align}

We are interested in estimating this expression
\begin{align}
    W_2  = \int_0^{\tau}~dt~\dfrac{\dot \lambda}{2} \bigg[\underbrace{\bigg(\int_{-\infty}^{+\infty}~dx_0~x_0^2 ~P_{\rm ss}(x_0)\bigg)}_{\sigma_{\rm ss}^2} G(t,0) + 2 D \int_0^t~ds~G(t,s)\bigg]\ , \label{w2avg}
\end{align}
where
\begin{align}
 \sigma_{\rm ss}^2 = Dt_U \zeta^2\ .   
\end{align}
and $\zeta$ is the dimensionless length~\eqref{app:eq:len}. $\zeta$ depends on both potentials as well as the distribution of times $f(t)$ in the exploration phase.
For example, in the case of an exponential distribution $f(t)= r e^{-rt}$ corresponding to a Poissonian resetting rate $r$, 

\begin{align}
 \zeta^2 = \left(\frac{rt_V+2}{rt_U + 2} \right)\ .   
\end{align}
and
\begin{align}
 \sigma_{\rm ss}^2(r) = Dt_U \left(\frac{rt_V+2}{rt_U + 2} \right)\ .   
\end{align}

We will now get both  the limiting value as well as the first-order correction as $\tau \rightarrow \infty$. We use the method sketched in \cite{Kwon_2013} to expand $W_2$ to first order in  $\alpha$.  To do this, note that we can write $W_2$ as
\begin{align}
    W_2  = \int_0^{\tau}~dt~\dfrac{\dot \lambda}{2} G(t,0) \bigg[\underbrace{\bigg(\int~dx_0~x_0^2 ~P_{\rm ss}(x_0)\bigg)}_{\sigma_{\rm ss}^2} + 2 D \int_0^t~ds~g(s,0)\bigg]\ , \label{w2avg_mod}
\end{align}
where 

  \begin{align}
    G(t,0)\equiv e^{-2\int_{0}^t~dt' \frac{\lambda(t')}{\gamma}} = e^{-(2\lambda_U/\gamma) (t + \alpha t^2 /2) }\ ,
\end{align}  
and 
\begin{align}
    g(s,0)\equiv e^{2\int_{0}^s~dt' \frac{\lambda(t')}{\gamma}} = e^{(2\lambda_U/\gamma) (s + \alpha s^2 /2) }\ .
\end{align}
where $\gamma = \frac{k_{\rm B}T}{D}$. Substituting the value of $\lambda(t)$ in Eq.~\eqref{w2avg_mod} gives us two terms:
\begin{align}
    W_2 = {\rm T}_1 + {\rm T}_2\,
\end{align}
where
\begin{align}
     {\rm T}_1&\equiv \dfrac{\alpha\lambda_U}{2}\sigma_{\rm ss}^2\int_0^{\tau}~dt~ G(t,0)\ ,\\
     {\rm T}_2&\equiv \alpha \lambda_U D\int_0^\tau~dt~  \int_0^t~ds~g(s,0)  \ .
\end{align}
We discuss first the following integral from T$_2$:
\begin{align}
  I \equiv D\lambda_U\alpha\int_0^t~ds~g(s,0)   \ .
\end{align}
We make a change of variables (as done in \cite{Kwon_2013})
\begin{align}
    u'= \frac{\lambda_U \alpha s}{\lambda_V - \lambda_U}\ ,
\end{align}
with this change of variables
\begin{align}
I = \lambda_U\alpha\int_0^t~ds~g(s,0) = D(\lambda_V - \lambda_U) \int_0^u du' e^{2bu' + c {u'}^2 }\ ,
\end{align}
where $u\equiv \frac{\lambda_U\alpha t}{\lambda_V-\lambda_U}$, $b = \frac{\lambda_V - \lambda_U}{\gamma \alpha}$
and $c= \frac{(\lambda_V - \lambda_U)^2}{\gamma \lambda_U \alpha} $.
\begin{align}
    I = D(\lambda_V - \lambda_U) \int_0^u~du'~e^{c ((2b/c) u' +  {u'}^2) }\ .
\end{align}
Note that both $b$ and $c$ diverge as $\alpha \rightarrow 0$. We will now perform an expansion of $O(\alpha)$.
We make a further change of variables $z= b/c + u' $, with this the integral of interest becomes
\begin{align}
   I = D (\lambda_V - \lambda_U) e^{-b^2/c}\int_{b/c}^{u+b/c} dz~ e^{cz^2 } \ .
\end{align}
We make a third change of variables $s=c z^2$ at which point the integral becomes
\begin{align}
 I \equiv D (\lambda_V - \lambda_U)  \frac{e^{-b^2/c}}{2\sqrt{c}}\int_{b^2/c}^{c(u+b/c)^2} ds \frac{e^{s}}{\sqrt{s}} \ .
\end{align}

We integrate this by parts to get
\begin{align}
I&= D (\lambda_V - \lambda_U) \frac{e^{-b^2/c}}{2\sqrt{c}}\left[\frac{e^{s}}{\sqrt{s}} + \frac{e^s}{2s^{3/2}} \right]\ ,\\
&=D (\lambda_V - \lambda_U) \frac{e^{-b^2/c}}{2\sqrt{c}}\frac{e^{s}}{\sqrt{s}}\bigg(1 + \frac{1}{2s}\bigg)\ ,
    \end{align}
to be evaluated at the limits of the integral.

\begin{align}
    I = D (\lambda_V - \lambda_U) \frac{e^{-b^2/c}}{2\sqrt{c}}\bigg[\frac{e^{c(u+b/c)^2}}{(u+b/c)\sqrt{c}}\bigg(1 + \frac{1}{2c(u+b/c)^2}\bigg) - \frac{e^{b^2/c}}{b/\sqrt{c}}\bigg(1 + \frac{1}{2b^2/c}\bigg)\bigg]\ ,\\
    =\frac{D (\lambda_V - \lambda_U) }{2c}\bigg[\frac{e^{cu^2+2bu}}{u+b/c}\bigg(1 + \frac{1}{2c(u+b/c)^2}\bigg) - \frac{1}{b/c}\bigg(1 + \frac{1}{2b^2/c}\bigg)\bigg]\ .
\end{align}

We hence have,
\begin{align}
   W_2= \int_0^{\tau} G(t,0) \left[ \frac{\lambda_U \alpha}{2}\sigma_{\rm ss}^2 + I(u(t)) \right] dt\ .
\end{align}

Making  a change of variables from $t$ to $u$,
\begin{align}
W_2&= \int_0^{1} G(u,0) \bigg[ \frac{\lambda_V - \lambda_U}{2}\sigma_{\rm ss}^2 +  \nonumber\\
 &+ \frac{D(\lambda_V-\lambda_U)^2}{2c\lambda_U\alpha} \bigg[\frac{e^{cu^2+2bu}}{u+b/c}\bigg(1 + \frac{1}{2c(u+b/c)^2}\bigg) - \frac{1}{b/c}\bigg(1 + \frac{1}{2b^2/c}\bigg)\bigg] du\ ,\\
 &= \int_0^{1} G(u,0) \bigg[ \frac{\lambda_V - \lambda_U}{2}\sigma_{\rm ss}^2 +  \nonumber\\
 &+ \dfrac{\gamma D}{2} \bigg[\frac{e^{cu^2+2bu}}{u+b/c}\bigg(1 + \frac{1}{2c(u+b/c)^2}\bigg) - \frac{1}{b/c}\bigg(1 + \frac{1}{2b^2/c}\bigg)\bigg] du\ .
    \end{align}

Let us evaluate the the following terms:
\begin{align}
    G(u,0) &= \exp\bigg[-\frac{2\lambda_U}{\gamma \alpha}\bigg(\xi u + \frac{\xi^2 u^2}{2}\bigg)\bigg]\ ,\\
    e^{cu^2+2bu} &= \exp\bigg[\frac{\lambda_U}{\gamma \alpha}\frac{(\lambda_V-\lambda_U)^2}{\lambda_U^2}u^2 + 2\dfrac{\lambda_U}{\gamma \alpha} \dfrac{\lambda_V-\lambda_U}{\lambda_U}u \bigg]=\exp\bigg[\frac{2\lambda_U}{\gamma \alpha}\bigg(\dfrac{\xi^2 u^2}{2} + \xi u\bigg) \bigg]\ ,
\end{align}
where $\xi \equiv \frac{\lambda_V-\lambda_U}{\lambda_U}$.
Hence,

\begin{align}
W_2&= \int_0^{1} \bigg[  G(u,0) \frac{\lambda_V - \lambda_U}{2}\sigma_{\rm ss}^2 +  \nonumber\\
 &+ \dfrac{\gamma D}{2} \bigg[\frac{1}{u+b/c}\bigg(1 + \frac{1}{2c(u+b/c)^2}\bigg) - \frac{G(u,0)}{b/c}\bigg(1 + \frac{1}{2b^2/c}\bigg)\bigg] du\ ,\\
 &=\int_0^{1} du~G(u,0) \bigg[  \frac{\lambda_V - \lambda_U} {2}\sigma_{\rm ss}^2 - \frac{\gamma D}{2b/c}\bigg(1 + \frac{1}{2b^2/c}\bigg)\bigg] +  \nonumber\\
 &+\int_0^{1}~du~\dfrac{\gamma D}{2} \bigg[\frac{1}{u+b/c}\bigg(1 + \frac{1}{2c(u+b/c)^2}\bigg)\bigg]  du\ ,\\
 &= {\rm II} + {\rm III}\ .
    \end{align}

\begin{align}
    {\rm II} &=\sqrt{\frac{\pi\alpha  \gamma }{4 \xi^2\lambda_U}}~e^{\frac{\lambda_U}{\alpha  \gamma }}  \left[\text{Erf}\left(\frac{\xi +1}{\sqrt{\frac{\alpha  \gamma }{\lambda_U}}}\right)-\text{Erf}\left(\frac{1}{\sqrt{\frac{\alpha  \gamma }{\lambda_U}}}\right)\right] \bigg[  \frac{\lambda_V - \lambda_U} {2}\sigma_{\rm ss}^2 - \frac{\gamma D}{2b}\bigg(1 + \frac{1}{2b^2/c}\bigg)\bigg]\ ,  \label{eq:II_1}\\
    &\approx \sqrt{\frac{\pi\alpha  \gamma }{4 \xi^2\lambda_U}}\bigg[\frac{\sqrt{\alpha } \sqrt{\frac{\gamma }{\lambda_U}}}{\sqrt{\pi }}-\frac{\sqrt{\alpha } \sqrt{\frac{\gamma }{\lambda_U}} e^{\frac{-\xi^2\lambda_U-2 \xi  \lambda_U}{\alpha  \gamma }}}{\sqrt{\pi } (\xi +1)}\bigg]\bigg[  \frac{\lambda_V - \lambda_U} {2}\sigma_{\rm ss}^2 - \frac{\gamma D}{2b/c}\bigg(1 + \frac{1}{2b^2/c}\bigg)\bigg]\ ,   \label{eq:II_2}\\
    &\approx \dfrac{\alpha \gamma}{2\xi \lambda_U}\bigg[\frac{\lambda_V - \lambda_U} {2}\sigma_{\rm ss}^2 - \dfrac{k_{\rm B}T}{2}\dfrac{\lambda_V-\lambda_U}{\lambda_U}\bigg]\ ,\\
\end{align}
where going from Eq.~\eqref{eq:II_1} to Eq.~\eqref{eq:II_2} we used Eqs.~\eqref{IG-1}--\eqref{IG-2}. 

Now, we evaluate III:
\begin{align}
    {\rm III} &= \dfrac{\gamma D}{2} \int_0^{1}~du~\bigg[\frac{1}{u+b/c}\bigg(1 + \frac{1}{2c(u+b/c)^2}\bigg)  du\ ,\\
    &= \dfrac{k_{\rm B}T}{2}\bigg[\ln \left(\frac{\lambda_V}{\lambda_U}\right)+\frac{\alpha  \gamma  \left(1-\frac{\lambda_U^2}{\lambda_V^2}\right)}{4 \lambda_U}\bigg]\ .
\end{align}

Putting these two integrals together the $\mathcal{O}(\alpha^0)$ term  becomes
\begin{align}
    W_2 [\mathcal{O}(\alpha^0)] = \dfrac{k_{\rm B}T}{2}\ln \left(\frac{\lambda_V}{\lambda_U}\right) = \dfrac{k_{\rm B}T}{2}\ln \left(\frac{t_U}{t_V}\right)\ .
    \label{eq:orderzero}
\end{align}

The term $\mathcal{O}(\alpha^1)$ is
\begin{align}
    W_2 [\mathcal{O}(\alpha^1)] &= \dfrac{\gamma}{4}\sigma_{\rm ss}^2 + \dfrac{k_{\rm B}T}{2}\frac{\gamma  \left(1-\frac{\lambda_U^2}{\lambda_V^2}\right)}{4 \lambda_U}- \dfrac{k_{\rm B}T\gamma}{4\lambda_U}\ ,\\
    &=\dfrac{k_{\rm B}T}{4} t_U \zeta^2 + \dfrac{k_{\rm B}T}{8}t_U\left(1-\frac{t_V^2}{t_U^2}\right) - \dfrac{k_{\rm B}Tt_U}{4}\ , \\
    &=\dfrac{k_{\rm B}T}{4} t_U\bigg[\zeta^2 -\dfrac{1}{2}\left(1+\frac{t_V^2}{t_U^2}\right)\bigg]\ .
\end{align}

As an example, for $f(t)= r e^{-rt}$, this term becomes
\begin{align}
    W_2 [\mathcal{O}(\alpha^1)] &= \dfrac{\gamma}{4}\sigma_{\rm ss}^2 + \dfrac{k_{\rm B}T}{2}\frac{\gamma  \left(1-\frac{\lambda_U^2}{\lambda_V^2}\right)}{4 \lambda_U}- \dfrac{k_{\rm B}T\gamma}{4\lambda_U}\ ,\\
    &=\dfrac{k_{\rm B}T}{4} t_U \dfrac{2+rt_V}{2+r t_U} + \dfrac{k_{\rm B}T}{8}t_U\left(1-\frac{r^2t_V^2}{r^2\lambda_U^2}\right) - \dfrac{k_{\rm B}Tt_U}{4}\ , \\
    &=\dfrac{k_{\rm B}T}{4} t_U\bigg[\dfrac{2+rt_V}{2+r t_U} -\dfrac{1}{2}\left(1+\frac{r^2t_V^2}{r^2 t_U^2}\right)\bigg] \ .\label{eq:corr}
\end{align}
Equation~\eqref{eq:corr} multiplied by $\alpha$ is the correction to the equilibrium free energy cost~\eqref{eq:orderzero} at late times.
An important point where this result deviates from the case studied in \cite{Kwon_2013}  is that the sign of the correction term  Eq.~\eqref{eq:orderzero} can be negative or positive. This implies that shorter protocols could have a lower or a higher work cost depending on the parameters. In contrast, in \cite{Kwon_2013}, the sign of the correction term is always positive implying that 
shorter protocol durations  always cost more. This is because in \cite{Kwon_2013},
the protocol operates between the equilibrium  states in the two traps, unlike the case we study here.

Finally, we provide some details of the steps involved in approximating the ${\rm Erf}$ function in Eq.~\eqref{eq:II_1}
to get Eq.~\eqref{eq:II_2}.
\begin{align}
    {\rm IG} \equiv \int_0^1~du~G(u,0)&= \int_0^1~du~\exp\bigg[-\frac{\lambda_U}{\gamma \alpha}\bigg(2\xi u + \xi^2 u^2\bigg)\bigg]\ , \label{IG-1}\\
    &=\int_0^1~du~\exp\bigg[-\frac{\lambda_U\xi^2}{\gamma \alpha}\bigg(2u/\xi + u^2\bigg)\bigg]\\
    &=e^{\lambda_U/(\gamma \alpha)}\int_0^1~du~\exp\bigg[-\frac{\lambda_U\xi^2}{\gamma \alpha}\bigg(1/\xi + u\bigg)^2\bigg]\ .
\end{align}

Substituting $\frac{\lambda_U\xi^2}{\gamma \alpha}\bigg(1/\xi + u\bigg)^2 = z$, then we have 
\begin{align}
    &2\frac{\lambda_U\xi^2}{\gamma \alpha}\bigg(1/\xi + u\bigg) du = dz \implies du = \dfrac{dz}{2\frac{\lambda_U\xi^2}{\gamma \alpha}\bigg(1/\xi + u\bigg)}\\
    &=\dfrac{dz}{2\sqrt{z} \sqrt{\frac{\lambda_U\xi^2}{\gamma \alpha}}  }\ .
\end{align}

\begin{align}
    {\rm IG} &=
    \dfrac{\sqrt{\gamma \alpha}}{2\xi \sqrt{\lambda_U}}e^{\lambda_U/(\gamma \alpha)}
    \int_{\frac{\lambda_U}{\gamma \alpha}}^{\frac{\lambda_U}{\gamma \alpha}(1 + \xi)^2}~du~\dfrac{e^{-z}}{\sqrt{z}}\ ,\\
    &=\dfrac{1}{2\xi}\dfrac{\gamma \alpha}{\lambda_U}\bigg[1 - \dfrac{e^{-\lambda_U/(\gamma \alpha)(\xi^2+2\xi)}}{1+\xi}\bigg] \ .\label{IG-2}
\end{align}
\subsection{Asymptotic value of the variance of the total work for the $\tanh$ protocol}
\label{var-asym}
{ The second moment of the total work has three contributions resulting from the second moment of $w_1$, the second moment of $w_2$  and the correlations between $w_1$ and $w_2$. These are given by the integrals $I_1$, $I_2$ and $I_3$ respectively, in Appendix~\ref{sec:sec_mom}. Subtracting out the average contribution from each, gives us the variance of each term, the sum being the variance of the total work. \\
{\bf Variance of $w_1$:}
Subtracting out the average contribution $W_1^2$ [Eq.~\eqref{w1avg}] from $I_1$, we immediately get the contribution of this term to the variance of the total work is $\dfrac{1}{2} D^2 t_V^2(\lambda_U-\lambda_V)^2\ $. It is this expression which is plotted as a dashed red line in the second panel of Fig.~\ref{fig:mean-var}, since, as we mention below, the other two terms contribute only time-dependent corrections. \\
{\bf Variance of $w_2$:}
{ It can be shown, using a similar method as detailed in this subsection~\ref{sec:long-time}, that this term contributes only
time-dependent terms to the variance of the total work in the long-time limit, similar to the analysis of \cite{Kwon_2013}. }
\\
{\bf The correlation term :}
{ This term is equal to $I_3 - \langle w_1\rangle\langle w_2\rangle$. Note that $I_3$ [Eq.~\eqref{soln-I3}] is very similar to the expression Eq.~\eqref{w2avg}. Hence, using the above analysis straightforwardly for 
$I_3$,
one can again show that this term too 
contributes only time-dependent correction terms to the asymptotic value of the variance of the total work.}
} 

\section{Proof of Eq.~\eqref{KLbound-main} }
\label{GLbound}
In this section, we provide the intermediate steps to obtain the GL work-bound mentioned in the main text. To this end, we follow Ref.~\cite{Esposito_2011}. 
The KL divergence between the time-dependent and the instantaneous equilibrium distribution is 
\begin{align}
    D_{\rm KL}[\rho(t)||\rho^{\rm eq}(t)] &\equiv \int~dx~\rho(x,t)\ln \dfrac{\rho(x,t)}{\rho^{\rm eq}(x,t)}\\
    &=\int~dx~\rho(x,t)\ln \dfrac{\rho(x,t)}{e^{-\beta H(t)+\beta F^{\rm eq}(t)}}\\
   & = \underbrace{\int~dx~\rho(x,t)\ln \rho(x,t)}_{-S_{\rm sys}/k_{\rm B}} + \beta[E(t) - F^{\rm eq}(t)]\ ,\label{KL-div}
\end{align}
where $E(t) \equiv \int~dx~\rho(x,t)H(t)$ is the average energy with respect to $\rho(t)$, and $F^{\rm eq}(t)$ is the instantaneous equilibrium free-energy at time $t$. We identify the first term on the right-hand side of Eq.~\eqref{KL-div} as the negative of the system entropy production, and we rewrite: 
\begin{align}
       \beta^{-1} D_{\rm KL}[\rho(t)||\rho^{\rm eq}(t)] 
       &=F(t) - F^{\rm eq}(t)\ , \label{this:eqn}
\end{align}
for the non-equilibrium free energy $F(t) = E(t) - T S_{\rm sys}$.

From the first law of thermodynamics, we have
\begin{align}
    W =  Q - \Delta E\ ,
\end{align}
where $\Delta E$, $Q$, and $W$, respectively, are the change in the internal energy of the system, heat absorbed by the system, and the work performed on the system. This equation can be rewritten as  
\begin{align}
     W &= -T [\Delta S_{\rm tot} - \Delta S_{\rm sys}] - \Delta E\ , 
\end{align}
for the change in total entropy production $\Delta S_{\rm tot}$.
We substitute $\Delta S_{\rm sys} \equiv S_{\rm sys}(t) - S_{\rm sys}(0)$, and exploit Eq.~\eqref{this:eqn} to obtain:
\begin{align}
   W -  \Delta F^{\rm eq} &=T \Delta S_{\rm tot} +k_{\rm B}T[D_{\rm KL}[\rho(t)||\rho^{\rm eq}(t)] -  D_{\rm KL}[\rho(0)||\rho^{\rm eq}(0)]]\ .
\end{align}

Since $\Delta S_{\rm tot}\geq 0$, we find the lower bound on the work performed while varying the control parameter $\lambda(t)$:
\begin{align}
    W \geq \Delta F^{\rm eq}  + k_{\rm B}T[D_{\rm KL}[\rho(t)||\rho^{\rm eq}(t)] -  D_{\rm KL}[\rho(0)||\rho^{\rm eq}(0)]]\ . \label{KLbound} 
\end{align}
For the control protocol $\lambda(t)$ changing from $\lambda_U\equiv \lambda(0)$ to $\lambda_V \equiv \lambda(\tau)$, we can simply translate the above expression to the average work $W_2$ as given in Eq.~\eqref{KLbound-main} by identifying that our system starts from $\rho(0) = P_{{\rm ta}, U}(x)$~\eqref{eq:noneqstate2} and the protocol finishes at $\rho(\tau) = P(x,\tau)$, and their corresponding instantaneous equilibrium probabilities, respectively, are $P^{\rm eq}_U(x)$ and $P^{\rm eq}_V(x)$.

\section{Derivation of the optimal protocol}\label{app:optimal}
Here we derive the details of the optimal protocols used in the main text. The optimal protocol connects the exploration phase, with trap stiffness $\lambda_U$, to the relaxation phase in the sharper trap with stiffness $\lambda_V$. When the protocol begins, the particle is in the state given by Eq.~\eqref{eq:noneqstate2}, which for brevity we repeat here as
\begin{equation}\label{eq:noneqstate_app}
    P_{{\rm ta}, U}(x_0) = \int_0^\infty~dt_1~f(t_1)~\int_{-\infty}^{+\infty}~dy~ P_{{\rm eq},V}(y)P_{U}(x_0,t_1|y)\ .
\end{equation}

The average work due to changing the stiffness during the protocol reads
\begin{equation}
    W_2[\{\lambda_t\}] = \int_0^\tau dt \left\langle \partial_{\lambda_t }\mathcal{U}(x_t,\lambda_t) \dot \lambda_t\right\rangle\ ,
\end{equation}
which reduces to the $n=1$ and $m=0$ terms on the right-hand side of Eq.~\eqref{eq:n-mom}. 
The optimal protocol $\lambda_t^{\rm opt}$ is the one that minimizes the work functional, $\frac{\delta W_2[\{\lambda_t^{\rm opt}\}]}{\delta_{\lambda_t^{\rm opt}} }   =0 $. The harmonic trap during the protocol phase is denoted
\begin{equation}
    \mathcal{U}(x_t,\lambda_t) = \frac{\lambda_t}{2}x_t^2\ ,
\end{equation}
and has a stiffness that is subject to the boundary conditions $\lambda_0 \equiv \lambda_U$ and $\lambda_\tau \equiv \lambda_V$. Here, the protocol duration is fixed to $\tau$, and $t \in (0,\tau)$ parametrizes the time-evolution throughout the protocol. Following Ref.~\cite{schmiedl2007optimal}, we can write the mean work as 
\begin{equation}\label{eq:w2opt}
    W_2 = \frac{1}{2} \left[\mathcal{V}_t \lambda_t - D\gamma \ln \mathcal{V}_t\right]_0^{\tau} + \frac{\gamma}{4} \int_0^{\tau} dt \frac{\dot {\mathcal{V}}_t^2}{\mathcal{V}_t}\ ,
\end{equation}
where $\mathcal{V}_\tau \equiv \langle x_\tau^2 \rangle$ is shorthand notation for the variance during the protocol, and $\gamma = (\beta D)^{-1}$ is the friction coefficient. The first term~\eqref{eq:w2opt} represents boundary contributions, while the integral in the second term can be interpreted as an action to be minimized with respect to the trajectory of $\mathcal{V}_t$. The Euler-Lagrange equation takes the form $[\dot {\mathcal{V}}_t]^2 = 2 \mathcal{V}_t \ddot {\mathcal{V}}_t$, with its solution 
\begin{align}\label{eq:var-sol}
\mathcal{V}_t =c_1(1+c_2t)^2\ .
\end{align} The two coefficients $c_i$ can be found through the boundary conditions of the protocol. We immediately see that $\mathcal{V}_0 = c_1$, and since at the beginning of the protocol the particle is in the state given by Eq.~\eqref{eq:noneqstate_app}, we have
\begin{align}
    c_1 &= \langle x^2 \rangle_\text{ta,U}=\int_0^\infty dt f(t)\int_{-\infty}^{+\infty} dx_0~ P_V^\text{eq}(x_0)~P_{U}(x,t|x_0)~x^2\\
   & =  \int_0^\infty dt f(t)\int_{-\infty}^{+\infty} dx_0~P^{\rm eq}_V(x_0) \left[x_0^2 e^{- 2t/t_U} + Dt_U (1-e^{-2t/t_U})\right]\\
   & = D t_U \zeta^2, \label{eq:app_c1}
\end{align}
where we substituted the dimensionless length $\zeta$ defined in Eq.~\eqref{eq:len}. The remaining constant, $c_2$, is found by directly minimizing the work 
\begin{equation}\label{eq:w2-optimal}
    W_2 = \frac{k_{\rm B}T}{2}\frac{t_U}{t_V} \zeta^2 (1+c_2 \tau)^2  - k_{\rm B}T \ln [1+c_2 \tau]-\frac{k_{\rm B}T}{2} \zeta^2  + k_{\rm B}T c_2^2 \tau  t_U \zeta^2\ . 
\end{equation}
with respect to $c_2$, i.e. by solving $\partial_{c_2} W_2(c_2) = 0$. This gives

\begin{equation}\label{eq:c2}
     c_2 = \frac{\sqrt{ \left(\frac{\tau}{\zeta^2 t_U} \left(2+\tau/t_V\right)+1\right)}- \left(1+\tau/t_V\right)}{\tau  \left( 2+ \tau/t_V\right)}\ .
\end{equation}
Substituting this $c_2$ in $W_2$~\eqref{eq:w2-optimal} gives $W_2^{\rm opt}$~\eqref{eq:work_opt_2} as reported in the main text.  Substituting $c_1$ and $c_2$ in Eq.~\eqref{eq:var-sol}, we obtain $\mathcal{V}_t^{\rm opt}$. 
To obtain the optimal protocol $\lambda_t^{\rm opt}$, we then use the equation of motion for the variance, $\dot {\mathcal{V}} = -2 D \beta \lambda \mathcal{V} + 2 D$, which ultimately gives
\begin{equation}\label{eq:appoptprot}
   \lambda_t^{\rm opt}  = \gamma\frac{  1 -  t_U \zeta^2 c_2 (1+c_2 t)}{  t_U \zeta^2 (1+c_2 t)^2}\ ,
\end{equation}
for $c_2$ given in Eq.~\eqref{eq:c2}.

In Fig.~\ref{fig:prot_app} we
study a few properties of the optimal protocol when the exploration time is chosen from a distribution $f(t)= r e^{-rt}$. Fig.~\ref{fig:prot_app}a shows the work associated with the optimal protocol  on fixing the relaxation time of the exploration trap $r t_U$ and varying $r t_V$ as well as the protocol duration $r \tau$. We see that the work is in general smaller than the equilibrium free energy difference only when the protocol is sufficiently long. Fig.~\ref{fig:prot_app}b shows one realization of the optimal protocol, showing the characteristic discontinuous jumps at the beginning and the end.

{  These can be understood better by considering the quantities $\Delta \lambda_{i,f}$ at the initial and final time of the protocol. The initial jump $\Delta \lambda_i = \lambda^\text{opt}_0 -\lambda_U$ can be calculated directly from Eq. (\ref{eq:appoptprot}). While the exact expression, which involves also $c_2$, is somewhat cumbersome, in the slow protocol limit it reads
\begin{equation}
    \lim_{\tau \to \infty} \Delta \lambda_i = \frac{\gamma}{t_U}\frac{1-\zeta^2}{\zeta^2}.
\end{equation}
From the bounds on the variable $\zeta$, i.e., $t_V/t_U\leq \zeta^2\leq 1$, we can infer that $ 0\leq \lim_{\tau \to \infty} \Delta \lambda_i\leq\lambda_V-\lambda_U$, meaning the jumps are always positive, and their largest value is exactly what is needed to jump all the way from $U$ to $V$. We recall that $\zeta$ is a direct measure of how far from equilibrium the initial state is, with $\zeta^2=1$ being the equilibrium case. Hence, the initial jump is a direct consequence of the non-equilibrium initial state. From  Eq. (\ref{eq:appoptprot}) one can also show that for the final jump $\Delta \lambda_f =\lambda_V-\lambda_\tau^\text{opt}$, the leading behavior for slow protocols scale as $\Delta \lambda_f \sim 1/\tau$, which vanishes for infinitely slow protocols. Fig.~\ref{fig:prot_app}c plots the size of these jumps as a function of protocol duration.
}

%

\begin{figure}
    \centering
    \includegraphics[width=17.6cm]{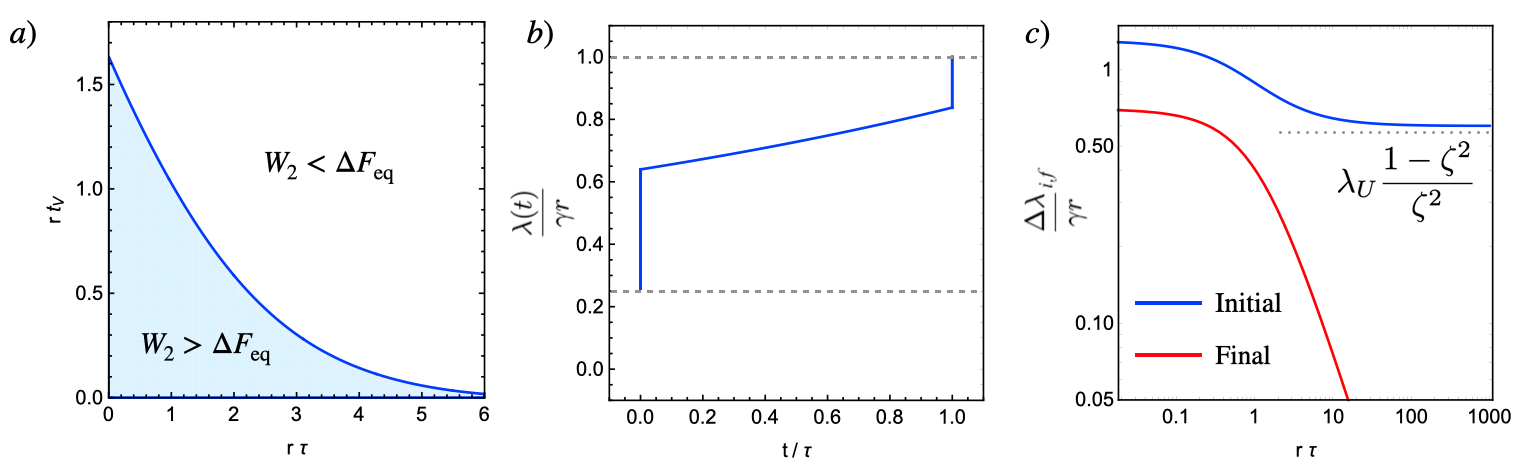}
    \caption{a) For finite protocol durations, the optimal protocol has regions where the associated work is higher or smaller than the equilibrium free energy difference. We have used $f(t) = r e^{-rt}$. For slow protocols, the work is always less than the equilibrium free energy, as discussed in the main text. Here $ r t_U = 3$. b) Example optimal protocol, for $r t_U = 4, r t_V = 1, r\tau = 1$.  c) The optimal protocol has discontinuous jumps at its initial and final stages, { with $\Delta \lambda_{i,f}$ described in the text}. While the initial jump discontinuity remains finite and approaches a constant as protocol duration is increased (dot-dashed line), the final jump goes to zero. This remaining asymmetry of the jumps in the quasistatic limit is a signature of the non-equilibrium initial state. Indeed, as $\tau \to \infty$ the initial jump is exactly the jump needed to match the variance of the initial state [given by Eq.~\eqref{eq:app_c1}]. Here, $r t_U= 4$ and $r t_V = 0.5$. }
    \label{fig:prot_app}
\end{figure}

We can separate the time-independent and time-dependent terms in Eq.~\eqref{eq:w2-optimal} using Eq.~\eqref{eq:c2} to get:
\begin{align}\label{eq:w2-optimal1}
    W_2 = &\Delta F_{U\to V}^\text{eq} + \frac{k_{\rm B} T}{2} \left[\ln\zeta^2 + (1-\zeta^2)\right]\ 
    -{k_{\rm B}T} \ln \left[ \frac{\sqrt{1+\frac{2t_V}{\tau} + \frac{\zeta^2 t_Ut_V}{\tau^2}}}{(1+\frac{2t_V}{\tau})} +\frac{\sqrt{\zeta^2 t_Ut_V}}{\tau (1+\frac{2t_V}{\tau})}\right] \nonumber \\ -& k_{\rm B}T \frac{t_U \zeta^2}{\tau }\left[ \frac{1}{\zeta} \sqrt{\frac{t_V}{t_U}} \frac{\sqrt{1+\frac{2t_V}{\tau} + \frac{\zeta^2 t_Ut_V}{\tau^2}}}{(1+\frac{2t_V}{\tau})} - \frac{(1+ \frac{t_V}{\tau})}{(1+ \frac{2t_V}{\tau})}\right]\ .
\end{align}

We expect the terms dependent on $\tau$ in  Eq.~\eqref{eq:w2-optimal1} to be strictly positive. This is easy to see if we carry out a large-$\tau$ expansion on the time-dependent terms. In this case  $W_2$ can be written as
\begin{equation}
  W_2 = \Delta F_{U\to V}^\text{eq} + \frac{k_{\rm B} T}{2} \left[\ln\zeta^2 + (1-\zeta^2)\right]\  + \frac{k_{\rm B}T}{2\tau} \left [ \sqrt{2t_V} - \zeta \sqrt{2t_U} \right]^2\ .
\end{equation}

Interestingly, the $\tau$-dependent term can be written as the square of a $L^2$- Wasserstein distance between two Gaussians, where one of the Gaussians is simply the Boltzmann distribution in the $V$ trap, while the other is a Gaussian in a modified shallow trap with variance $D\zeta^2 t_U$. The out-of-equilibrium nature of the initial state is characterized by $\zeta$ with  $t_V/t_U \leq \zeta^2 \leq  1$. 

To do the small- $\tau$ expansion, multiply and divide the terms in both square brackets in Eq.~\eqref{eq:w2-optimal1}  by $\tau$ and expand upto $\mathcal{O}(\tau^1)$. The first term gives a $\tau$-independent contribution to highest order while the highest order contribution of the second term is $\mathcal{O}(\tau^1)$. This recovers Eq. (\ref{eqn:deltaWa})
in the main text.
We are therefore able to characterize the full time dependence of the optimal work cost $W_2$, as well as the long-time and short-time limits.

.

\section{Calculation of $A(t_1,t_2)$}
\label{sec:Aw2}
Let us evaluate the following average:
\begin{align}
     A(t_1,t_2) = \int_{-\infty}^{+\infty}~dx_0~P_{{\rm ta},U}(x_0)\underbrace{\langle x^2(t_1) x^2(t_2)\rangle_{0,x_0}}_{a(t_1,t_2,x_0)}\ ,\label{eq:A-int}
\end{align}
where $P_{{\rm ta},U}(x_0)$ is the time-average density~\eqref{app:ta-U-den}, and $x(t)$ is the particle's position at time $t$ in a time-dependent harmonic potential $\mathcal{U}(x;\lambda(t))$ such that the stiffness $\lambda$  changes from $\lambda_U$ to $\lambda_V$. Each trajectory is initialized from $x_0$ drawn from the time-average distribution $P_{{\rm ta},U}(x_0)$.

For a time-dependent harmonic potential, the equation of motion reads
\begin{align}
    \dot x = -\beta D\lambda(t)x+ \sqrt{2D}~\eta(t)\ .
\end{align}
The solution of the above equation is 
\begin{align}
    x(t) = x_0 g(t,0) +  \sqrt{2D} \int_0^t~ds~g(t,s)~\eta(s)\ , \label{soln-x0}
\end{align}
where the first term on the right-hand side is the average position of the particle in the time-varying potential when the particle starts from $x_0$ drawn from $P_{{\rm ta},U}(x_0)$~\eqref{app:ta-U-den}. In the above equation~\eqref{soln-x0}, we have:
\begin{align}
    g(t,s)\equiv e^{-\beta D\int_{s}^t~dt' \lambda(t')}\ .
\end{align}

We can rewrite $x(t)$~\eqref{soln-x0} as  
\begin{align}
    x(t) = \mu(t) + y(t)\ ,
\end{align}
where we have defined 
\begin{align}
    \mu(t) &\equiv x_0 g(t,0)\ ,\\
    y(t)&\equiv \sqrt{2D} \int_0^t~ds~g(t,s)~\eta(s)\ ,
\end{align}
where $y(t)$ is a normal random variable with zero mean.

Then, $a(t_1,t_2,x_0)$ in Eq.~\eqref{eq:A-int} becomes
\begin{align}
     a(t_1,t_2,x_0) &= \langle x^2(t_1) x^2(t_2)\rangle_{0,x_0}\\
     &=\langle [y^2(t_1)+\mu^2(t_1) + 2 y(t_1)\mu(t_1)][y^2(t_2)+\mu^2(t_2) + 2 y(t_2)\mu(t_2)]\rangle_{0,x_0}\ .
\end{align}

Expanding the terms on the right-hand side of Eq.~\eqref{eq:A-int}, and then, performing an average over the time-average distribution $P_{{\rm ta},U}(x_0)$~\eqref{app:ta-U-den}, we get
\begin{align}
    A(t_1,t_2) 
     &=\langle y^2(t_1)y^2(t_2) \rangle + 2\langle \mu^2(t_1) \rangle_{x_0}\langle y^2(t_2)\rangle + 4 \langle \mu(t_1)\mu(t_2) \rangle_{x_0} \langle y(t_1)y(t_2) \rangle + \langle \mu^2(t_1)\mu^2(t_2) \rangle_{x_0} \ , \label{A-rhs}
\end{align}
where the $\langle [\cdots] \rangle_{x_0} \equiv \int~dx_0~P_{{\rm ta},U}(x_0) [\cdots]_{0,x_0}$ [see Eq.~\eqref{eq:A-int}].

Now, we use Wick's theorem for Gaussian random variables:
\begin{align}
       \langle y^2(t_1) y^2(t_2)\rangle = \langle y^2(t_1)\rangle\langle y^2(t_2)\rangle + 2\langle y(t_1) y(t_2)\rangle^2 \ ,
\end{align}
and rewrite the right-hand side of Eq.~\eqref{A-rhs} to get
\begin{align}
     A(t_1,t_2) &=\langle y^2(t_1)\rangle\langle y^2(t_2)\rangle + 2\langle y(t_1) y(t_2)\rangle^2+ 2\langle \mu^2(t_1) \rangle_{x_0}\langle y^2(t_2)\rangle \nonumber\\
     &+ 4 \langle \mu(t_1)\mu(t_2) \rangle_{x_0} \langle y(t_1)y(t_2) \rangle + \langle \mu^2(t_1)\mu^2(t_2) \rangle_{x_0}\ . \label{eq:A}
\end{align}

The right-hand side of the above equation~\eqref{eq:A} can be calculated using the following integrals:
\begin{align}
    \langle y^2(t_i)\rangle&=2D\int_0^{t_i}~ds~G(t_i,s)\ ,\\
    \langle y(t_1) y(t_2)\rangle&=2D\int_0^{{\rm min}(t_1,t_2)}~ds~g(t_1,s)g(t_2,s)\ ,\\
    \langle \mu^2(t_i) \rangle_{x_0} &= \bigg(\int_{-\infty}^{+\infty}~dx_0~x_0^2 P_{{\rm ta},U}(x_0)\bigg) G(t_i,0)\ ,\\
    \langle \mu(t_1)\mu(t_2) \rangle_{x_0} & = \bigg(\int_{-\infty}^{+\infty}~dx_0~x_0^2 P_{{\rm ta},U}(x_0)\bigg) g(t_1,0)g(t_2,0)\ ,\\
    \langle \mu^2(t_1)\mu^2(t_2) \rangle_{x_0} & = \bigg(\int_{-\infty}^{+\infty}~dx_0~x_0^4 P_{{\rm ta},U}(x_0)\bigg) G(t_1,0)G(t_2,0)\ .
\end{align}

\end{widetext}


%

\end{document}